\documentclass[intlimits,twoside,a4paper]{article}

\usepackage{amsmath,amssymb}
\usepackage{braket}
\usepackage{color}
\usepackage{ulem}

\usepackage[eqsecnum]{cmpj3}

\usepackage{bm}

\issue{2026}{29}{3}{33705}
\doinumber{10.5488/CMP.29.33705}

\title[Competition between uniform ferromagnetism and finite-wave-vector instabilities]%
{Competition between uniform ferromagnetism and finite-wave-vector instabilities in square-lattice skyrmion crystals}
\author[S. Hayami]{S. Hayami\orcid{0000-0001-9186-6958}\thanks{Email: \email{hayami@phys.sci.hokudai.ac.jp}.},
}
\address{
Graduate School of Science, Hokkaido University, Sapporo 060-0810, Japan
}


\date{Received 15 June 2026; revised 22 July 2026; accepted 16 August 2026; published 28 September 2026}

\begin{document}

\maketitle

\begin{abstract}
	 We investigate square-lattice skyrmion crystal states in a centrosymmetric spin model with momentum-depen\-dent magnetic interactions, including a uniform ferromagnetic interaction, easy-axis anisotropic interactions at two orthogonal finite wave vectors, higher-harmonic wave-vector interactions, and an external magnetic field. Using simulated annealing calculations, we determine the low-temperature magnetic phase diagram as functions of magnetic field and uniform ferromagnetic interaction strength. We find that the skyrmion crystal phase is stabilized at intermediate fields when the uniform ferromagnetic interaction is weak, exhibiting a noncoplanar double-$Q$ spin texture, finite scalar spin chirality, and higher-harmonic components. Increasing the uniform ferromagnetic interaction suppresses and eventually eliminates the skyrmion crystal phase through direct first-order transitions to the fully polarized state, whereas its weakening  expands the skyrmion stability region and stabilizes a single-$Q$ conical spiral phase at high fields. These results clarify the mechanisms governing the square-lattice skyrmion formation in centrosymmetric magnets.

\printkeywords
\end{abstract}

\section{Introduction}

Skyrmion crystals (SkXs) have attracted considerable attention in condensed matter physics because of their topologically nontrivial spin textures and their potential applications in spintronic devices~\cite{nagaosa2013topological, Tokura_doi:10.1021/acs.chemrev.0c00297, gobel2021beyond, guslienko2024magnetic}. 
In particular, the particle-like nature, nanoscale size, and efficient current-driven dynamics of skyrmions provide promising routes toward next-generation information-storage and logic technologies~\cite{Jonietz_skyrmion, yu2012skyrmion, fert2013skyrmions, zhang2015magnetic, fert2017magnetic, jiang2017direct, Baltz_RevModPhys.90.015005, luo2018reconfigurable, Chauwin_PhysRevApplied.12.064053, zhang2020skyrmion, yu2020motion}. 
While early studies mainly focused on SkX states stabilized by the Dzyaloshinskii--Moriya (DM) interaction~\cite{dzyaloshinsky1958thermodynamic,moriya1960anisotropic} in noncentrosymmetric magnets~\cite{rossler2006spontaneous, Binz_PhysRevLett.96.207202, Binz_PhysRevB.74.214408, Yi_PhysRevB.80.054416, Butenko_PhysRevB.82.052403, heinze2011spontaneous, Wilson_PhysRevB.89.094411}, such as MnSi~\cite{Muhlbauer_2009skyrmion, Neubauer_PhysRevLett.102.186602, Adams_PhysRevLett.107.217206, Morikawa_PhysRevB.88.024408, Bauer_PhysRevLett.110.177207, Wilson_PhysRevB.89.094411}, Fe$_{1-x}$Co$_x$Si~\cite{yu2010real, adams2010skyrmion, Munzer_PhysRevB.81.041203, Bauer_PhysRevB.93.235144}, Cu$_2$OSeO$_3$~\cite{chacon2018observation,takagi2020particle}, and EuPtSi~\cite{kakihana2018giant,kaneko2019unique,tabata2019magnetic, kakihana2019unique, Mishra_PhysRevB.100.125113, takeuchi2019magnetic, hayami2021field, Matsumura_PhysRevB.109.174437}, subsequent theoretical and experimental developments have revealed that noncoplanar multiple-$Q$ magnetic states can also emerge in centrosymmetric systems through competing exchange interactions and/or itinerant-electron effects~\cite{Okubo_PhysRevLett.108.017206, leonov2015multiply, Hayami_PhysRevB.94.174420, Hayami_PhysRevB.103.224418, Hayami_PhysRevB.105.014408, lin2024skyrmion, hayami2024stabilization, kawamura2025frustration}. 
Representative examples include Gd$_2$PdSi$_3$~\cite{Saha_PhysRevB.60.12162, kurumaji2019skyrmion, sampathkumaran2019report, Hirschberger_PhysRevB.101.220401, Kumar_PhysRevB.101.144440, Spachmann_PhysRevB.103.184424, Gomilsek_PhysRevLett.134.046702}, Gd$_3$Ru$_4$Al$_{12}$~\cite{chandragiri2016magnetic, Nakamura_PhysRevB.98.054410, hirschberger2019skyrmion, Nakamura_PhysRevB.111.184433}, and GdRu$_2$Si$_2$~\cite{khanh2020nanometric, Wood_PhysRevB.107.L180402, eremeev2023insight, Huddart_PhysRevB.111.054440, dong2025pseudogap}, where field-induced SkX phases have been experimentally observed despite the absence of inversion-symmetry breaking. 
These findings have significantly broadened the class of materials capable of hosting topological spin textures and have stimulated intensive studies of skyrmion formation mechanisms beyond chiral magnets~\cite{Tokura_doi:10.1021/acs.chemrev.0c00297}.

A remarkable feature of centrosymmetric skyrmion systems is that the stabilization mechanism is often governed by the competition among several magnetic instabilities characterized by distinct ordering wave vectors related by the crystal symmetry. 
In itinerant magnets, such instabilities naturally arise from the momentum dependence of effective spin interactions mediated by conduction electrons~\cite{hayami2024stabilization}. 
In particular, finite-$Q$ interactions originating from Fermi-surface effects favor spatially modulated magnetic structures, while the coexistence of different ordering wave vectors can generate noncoplanar multiple-$Q$ states through nonlinear mode coupling. 
This mechanism provides a unified framework for understanding a variety of exotic magnetic textures, including SkX states~\cite{Wang_PhysRevLett.124.207201, Bouaziz_PhysRevLett.128.157206, wang2023skyrmion}, vortex crystals~\cite{Solenov_PhysRevLett.108.096403, takagi2018multiple, park2025spin}, hedgehog crystals~\cite{Eto_PhysRevLett.132.226705, Kato_PhysRevB.104.224405, Okumura_doi:10.7566/JPSJ.91.093702}, and meron-antimeron crystals~\cite{Hayami_PhysRevB.104.094425, Mohylna_PhysRevB.111.174435}, realized in centrosymmetric magnets.

Among these states, square-lattice SkX states have recently attracted increasing attention because their stabilization mechanism qualitatively differs from that of conventional triangular-lattice SkX states. 
In triangular-lattice systems, triple-$Q$ superpositions naturally emerge owing to the symmetry equivalence among three ordering wave vectors related by lattice rotations. 
By contrast, square-lattice SkX states are typically associated with the superposition of two orthogonal magnetic modulations. 
Such square-type SkX states have been discussed particularly in tetragonal itinerant magnets such as GdRu$_2$Si$_2$~\cite{khanh2020nanometric, Wood_PhysRevB.107.L180402, eremeev2023insight, Huddart_PhysRevB.111.054440, dong2025pseudogap}, where magnetic anisotropy and competing interactions strongly influence the stability of double-$Q$ magnetic structures. 
As a consequence, the parameter region supporting square-lattice SkX states is often relatively narrow, and small perturbations can significantly modify the magnetic phase competition~\cite{Christensen_PhysRevX.8.041022, Hayami_doi:10.7566/JPSJ.89.103702, Utesov_PhysRevB.103.064414, Wang_PhysRevB.103.104408, Hayami_PhysRevB.105.174437, Hayami_PhysRevB.105.104428, utesov2025thermodynamic}.

Recent theoretical studies have suggested that magnetic anisotropies play an essential role in stabilizing square-type SkX textures in centrosymmetric systems~\cite{Utesov_PhysRevB.103.064414, Wang_PhysRevB.103.104408, Hayami_PhysRevB.105.174437}. 
Easy-axis anisotropies, in particular, tend to enhance the out-of-plane spin component and promote noncoplanar double-$Q$ states composed of orthogonal modulations. 
At the same time, high-harmonic wave-vector interactions can stabilize multiple-$Q$ magnetic states even in the absence of explicit DM interactions~\cite{Hayami_PhysRevB.105.174437, hayami2022multiple}. 
Such high-harmonic wave-vector interactions are especially important in frustrated magnets because they provide nonlinear couplings among symmetry-related ordering wave vectors and strongly influence the selection of magnetic textures.

Despite these advances, the interplay between uniform ferromagnetic interactions and finite-$Q$ anisotropic interactions has not yet been systematically clarified within momentum-space spin models~\cite{hayami2024stabilization}. 
In many centrosymmetric systems, the magnetic energy contains both a $\bm{q}=\bm{0}$ uniform component favoring collinear ferromagnetism and finite-$q$ interactions favoring spatially modulated spin textures. 
The competition between these two tendencies is expected to play a decisive role in determining the stability of noncoplanar multiple-$Q$ states. 
However, previous studies have mainly focused either on finite-$Q$ instabilities alone or on phenomenological spin models without explicitly examining how the balance between uniform and finite-wave-vector interactions modifies the stability region of square-lattice SkX states.

Motivated by these considerations, in the present study, we investigate the stability of square-lattice SkX states in centrosymmetric magnets by considering a spin model on a square lattice with momentum-dependent magnetic interactions in reciprocal space. 
In particular, we focus on the competition between the uniform ($\bm{q}=\bm{0}$) ferromagnetic interaction and anisotropic finite-$Q$ interactions that preferentially enhance the spin component along the $z$ direction and promote modulated magnetic states with easy-axis character. 
We also incorporate higher-harmonic wave-vector interactions that favor the superposition of orthogonal magnetic modulations and thereby enhance the tendency toward double-$Q$ ordering. 
To clarify the interplay among these interactions, we perform simulated annealing calculations and systematically determine the low-temperature magnetic phase diagram together with the corresponding spin textures. 
Our results demonstrate that the strength of the $\bm{q}=\bm{0}$ ferromagnetic interaction qualitatively changes the stability of the square-lattice SkX phase. 
For sufficiently large uniform interactions, the SkX phase disappears, and the system exhibits a first-order transition from either a single-$Q$ vertical spiral state or a single-$Q$ collinear state directly into the fully polarized state. 
On the other hand, when the $\bm{q}=\bm{0}$ interaction is reduced, the stability region of the SkX phase is significantly expanded in the intermediate-field region. 
In addition, a single-$Q$ conical spiral state emerges in the high-field region for weaker uniform interactions. 
These results reveal how the competition between uniform ferromagnetism, anisotropic finite-wave-vector instability, and higher-harmonic mode coupling controls the stability of square-lattice SkX states in centrosymmetric systems.

The rest of this paper is organized as follows. 
In section~\ref{sec: Model and method}, we introduce the centrosymmetric spin model on a square lattice with momentum-dependent magnetic interactions and explain the simulated annealing method used in the present study. 
Section~\ref{sec: Results} presents the low-temperature magnetic phase diagrams obtained for different strengths of the uniform ferromagnetic interaction and discusses the corresponding magnetic textures, spin structure factors, and field-induced phase transitions. 
We particularly focus on how the competition between the $\bm{q}=\bm{0}$ ferromagnetic interaction, anisotropic finite-$Q$ interactions, and higher-harmonic wave-vector interactions controls the stability of square-lattice SkX states in the presence of an external magnetic field. 
Finally, section~\ref{sec: Conclusions} is devoted to a summary of the present results.

\section{Model and method}
\label{sec: Model and method}

\subsection{Spin model with momentum-space interactions}

To investigate the stability of square-lattice SkX states in centrosymmetric magnets, we consider a classical spin model on a square lattice. 
The model consists of bilinear interactions specified in momentum space and is designed to describe the competition among the uniform ferromagnetic component, the fundamental finite-$Q$ modulations, and the higher-harmonic wave-vector components generated by their superpositions. 
The Hamiltonian is written as~\cite{Yambe_PhysRevB.106.174437}
\begin{align}
\mathcal{H}
= \mathcal{H}_{0} + \mathcal{H}_{Q} + \mathcal{H}_{Q_{\rm high}} + \mathcal{H}_{\rm Z},
\label{eq:Hamiltonian}
\end{align}
where each term is defined below.

The first term represents the uniform ferromagnetic interaction at $\bm q=\bm 0$:
\begin{align}
\mathcal{H}_{0}
=
-J_{0}
\sum_{\alpha=x,y,z}
(S_{\bm 0}^{\alpha})^2.
\label{eq:H0}
\end{align}
This term favors the uniform alignment of spins and competes with the finite-wave-vector interactions that stabilize spatially modulated magnetic states. 
Here, the Fourier component of the spin is defined as
\begin{align}
S_{\bm q}^{\alpha}
=
\frac{1}{\sqrt{N}}
\sum_i
S_\ri^{\alpha}
\re^{-\ri\bm q\cdot \bm r_i},
\end{align}
where $\bm S_i=(S_i^x,S_i^y,S_i^z)$ is a classical spin satisfying $|\bm S_i|=1$, $N$ is the number of lattice sites, and $\bm r_i$ is the position vector of site $i$.

The second term describes the fundamental finite-$Q$ interactions at two orthogonal ordering wave vectors,
\begin{align}
\bm Q_1=(Q,0), \qquad \bm Q_2=(0,Q),
\end{align}
with $Q=\piup/3$. 
These ordering wave vectors are responsible for single-$Q$ and double-$Q$ magnetic modulations on the square lattice. 
We introduce an easy-axis Ising-type anisotropy for these interactions as
\begin{align}
\mathcal{H}_{Q}
= -2\sum_{\nu=1,2}
\left[ J_{Q}^{\perp} \left( S_{\bm Q_\nu}^{x}S_{-\bm Q_\nu}^{x} +
S_{\bm Q_\nu}^{y}S_{-\bm Q_\nu}^{y} \right) + J_{Q}^{z} S_{\bm Q_\nu}^{z}S_{-\bm Q_\nu}^{z} \right],
\label{eq:HQ}
\end{align}
where $J_{Q}^{z}>J_{Q}^{\perp}$ favors out-of-plane spin modulations. 
This anisotropic finite-$Q$ interaction tends to promote noncoplanar magnetic structures, such as the SkX, in an external magnetic field by enhancing the $z$-spin component at the fundamental ordering wave vectors~\cite{hayami2020multiple}.

The remaining bilinear interactions are introduced at the higher-harmonic wave vectors generated from the two fundamental modulations,
\begin{align}
\bm Q_{3}=\bm Q_1+\bm Q_2, \qquad
\bm Q_{4}=\bm Q_1-\bm Q_2 .
\end{align}
Their contributions are given by
\begin{align}
\mathcal{H}_{Q_{\rm high}}
=
-2J_{Q_{\rm high}}
\sum_{\nu={3,4}}\sum_{\alpha=x,y,z}
S_{\bm Q_{\nu}}^{\alpha}S_{-\bm Q_{\nu}}^{\alpha}.
\label{eq:HQp}
\end{align}
These terms do not introduce explicit nonlinear interactions; instead, they represent additional bilinear exchange channels at the higher-harmonic wave vectors. 
They energetically support spin configurations containing Fourier components at $\bm Q_1\pm\bm Q_2$, which naturally arise when two orthogonal fundamental modulations coexist. 
Thus, the higher-harmonic wave-vector interactions help stabilize double-$Q$ magnetic states such as the square-lattice SkX~\cite{hayami2022multiple, hayami2023widely}.

The last term is the Zeeman coupling to an external magnetic field:
\begin{align}
\mathcal{H}_{\rm Z}
=
-H\sum_i S_i^{z}.
\label{eq:HH}
\end{align}
Throughout this study, the magnetic field is applied along the $z$ direction. 

In the present model, the central control parameter is the relative strength of the uniform ferromagnetic interaction $J_0$ against the finite-wave-vector interactions. 
A large $J_0$ favors a collinear ferromagnetic alignment and suppresses the development of noncoplanar multiple-$Q$ textures. 
By contrast, reducing $J_0$ enhances the relative importance of the finite-$Q$ and higher-harmonic wave-vector interactions, thereby expanding the stability region of square-lattice SkX states and allowing other modulated phases, such as single-$Q$ conical spirals, to appear. 
In the following calculations, we fix the interaction parameters as $J_{Q}^{\perp}=J-A/2$, $J_{Q}^{z}=J+A$, and $J_{Q_{\rm high}}=0.6$ with $A=0.3$, while taking $J=1$ as the energy unit. 
The strength of the uniform ferromagnetic interaction is parameterized as $J_0=\gamma J$, and the magnetic phase diagram is investigated by varying $\gamma$ and the external magnetic field $H$.

{
The uniform interaction at $\bm{q}=\bm{0}$ naturally arises as the zero-wave-vector component of the Fourier-transformed exchange interaction.
For example, starting from a generic real-space spin Hamiltonian,
\begin{align}
\mathcal{H}
=
-\sum_{ij}
J_{ij}
\,\bm{S}_{i}\cdot\bm{S}_{j},
\end{align}
its Fourier transformation yields
\begin{align}
\mathcal{H}
=
-\sum_{\bm{q}}
J(\bm{q})
\,
\bm{S}_{\bm{q}}
\cdot
\bm{S}_{-\bm{q}},
\end{align}
where $J(\bm{q}=\bm{0})$ corresponds to the uniform ferromagnetic interaction, while maxima at finite wave vectors describe competing magnetic instabilities toward modulated states.
Such momentum-dependent interactions naturally emerge from microscopic exchange mechanisms, including short-range exchange interactions in localized spin systems and long-range Ruderman--Kittel--Kasuya--Yosida interactions in itinerant magnets~\cite{Ruderman, Kasuya, Yosida1957}.
Within this effective momentum-space model, the parameter $\gamma$ is introduced as an independent tuning parameter that controls the relative strength of the $\bm{q}=\bm{0}$ interaction with respect to the finite-$q$ interactions, thereby enabling a systematic investigation of the competition between uniform ferromagnetism and finite-wave-vector magnetic instabilities.
}

Such momentum-space effective spin models provide an efficient framework for describing magnetic instabilities with specific spatial periodicities. 
Compared with real-space spin models, they enable direct access to the dominant ordering wave vectors without explicitly treating complicated long-range interactions in real space. 
In itinerant magnets, this approach effectively captures the magnetic instability arising from Ruderman-Kittel-Kasuya-Yosida interactions~\cite{Ruderman, Kasuya, Yosida1957} mediated by conduction electrons owing to the Fermi-surface instability~\cite{yoshimori1959new, Kaplan_PhysRev.124.329, Elliott_PhysRev.124.346}, while in frustrated localized-spin systems, it can phenomenologically incorporate the competition among multiple exchange interactions through the structure of momentum-dependent couplings~\cite{day1981neutron, regnault1982inelastic, nakatsuji2005spin}. 
For this reason, momentum-space effective models have been widely employed to investigate the emergence of topological magnetic phases, including multiple-$Q$ SkX states~\cite{yambe2021skyrmion}, as well as to explain complex magnetic phase diagrams observed in real materials like CeAuSb$_2$~\cite{seo2021spin}, EuNiGe$_3$~\cite{singh2023transition}, and EuAg$_4$Sb$_2$~\cite{neves2026cascade, neves2026general}.

\subsection{Numerical method}

The low-temperature magnetic phases of the Hamiltonian in equation~(\ref{eq:Hamiltonian}) are determined by simulated annealing based on the classical Metropolis Monte Carlo method. 
The simulations are performed without imposing any assumed magnetic structure, so that single-$Q$, double-$Q$, and fully polarized states can be compared on the same footing.

We use square lattices with periodic boundary conditions. 
The system size is chosen so that the ordering wave vectors $\bm Q_1$ and $\bm Q_2$ are commensurate with the lattice ($N=12^2$), which avoids artificial frustration from the boundary condition. 
{
To assess finite-size effects, we additionally performed simulations for a larger lattice with $L=24$.
The field dependence of the magnetization, scalar spin chirality, and Fourier amplitudes is nearly identical for $L=12$ and $L=24$, leading to the same magnetic phase sequence, phase boundaries, and spin textures.
Furthermore, no competing incommensurate states were found in the parameter range investigated.
}

Starting from random spin configurations at high temperature $T_0=1$--$3$, the temperature is gradually reduced to the low-temperature region according to $T_{n+1}=\tilde{\alpha}T_n$ with $\tilde{\alpha}=0.999995$--$0.999999$, where $T_n$ is the $n$th temperature. 
The final temperature is set to $T=10^{-4}$. 
At each temperature step, Monte Carlo updates are performed for equilibration, and the final spin configurations are used to identify the stable magnetic phases. 
Physical quantities are measured in the low-temperature regime by performing statistical averages over $10^5$--$10^6$ Monte Carlo sweeps after sufficient thermalization.
To reduce the possibility of trapping in metastable states, we repeat the calculations from several independent initial conditions and also check configurations obtained near phase boundaries.

The magnetic phases are characterized using real-space spin textures, spin structure factors, and scalar spin chirality. 
The spin structure factor for each spin component is defined by
\begin{align}
S_s^{\alpha}(\bm q)
=
\frac{1}{N}
\sum_{i,j}
S_i^{\alpha}S_j^{\alpha}
\re^{\ri\bm q\cdot(\bm r_i-\bm r_j)},
\qquad
\alpha=x,y,z .
\end{align}
The peak positions and relative intensities of $S_s^\alpha(\bm q)$ are used to distinguish single-$Q$ states from double-$Q$ SkX states and to identify the contribution of the higher-harmonic wave-vector components.
To quantify the contribution from each ordering wave vector, we also introduce the momentum-dependent magnetic amplitude defined by
\begin{align}
m^\eta_{\bm{q}}=\sqrt{\frac{S^\eta_s(\bm{q})}{N}}.
\end{align}
This quantity characterizes the relative intensity of magnetic modulations in momentum space and is useful for identifying the dominant ordering wave-vector components in competing magnetic states.
The in-plane contribution is extracted as $S_s^{xy}(\bm{q})=S_s^{x}(\bm{q})+S_s^{y}(\bm{q})$ and $m^{xy}_{\bm{q}}=\sqrt{S^{xy}_s(\bm{q})/N}$. 
In order to monitor the field-induced evolution of the magnetic states, we evaluate the magnetization parallel to the external magnetic field as
\begin{align}
M^z=\frac{1}{N}\sum_i S_i^z .
\end{align}

To characterize the degree of noncoplanarity in the magnetic textures, we calculate the scalar spin chirality defined on elementary plaquettes of the square lattice:
\begin{align}
\chi^{\mathrm{sc}} &= \frac{1}{N}\sum_i \sum_{\delta=\pm1} 
\bm{S}_i \cdot 
\left(
\bm{S}_{i+\delta\hat{x}} 
\times 
\bm{S}_{i+\delta\hat{y}}
\right) \nonumber \\
&\equiv \frac{1}{N}\sum_i \chi_i^{\mathrm{sc}},
\end{align}
where $\hat{x}$ and $\hat{y}$ are the unit lattice vectors along the horizontal and vertical directions, respectively. 
This quantity measures the solid angle formed by the neighboring spins and therefore serves as an indicator of topological magnetic textures. 
Nonzero values of $\chi^{\mathrm{sc}}$ signal the presence of SkX phases.

\section{Results}
\label{sec: Results}

\begin{figure}[ht!]
\begin{center}
\includegraphics[width=7.0 cm]{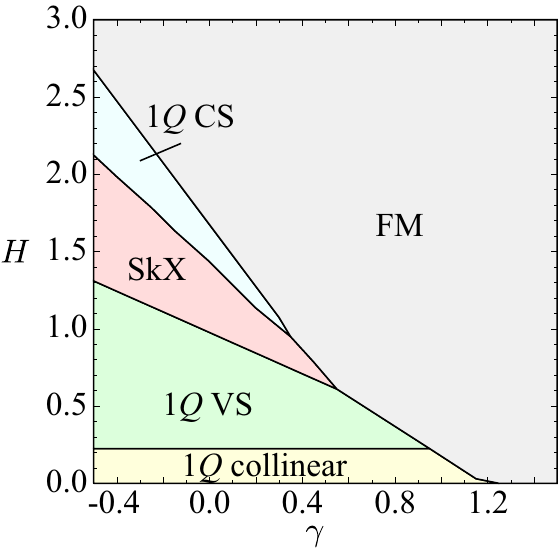}
\caption{
\label{fig: PD}
(Colour online) Low-temperature magnetic phase diagram in the $\gamma$--$H$ plane obtained by simulated annealing calculations for the momentum-space spin model in equation~(\ref{eq:Hamiltonian}). 
Here, $\gamma$ controls the relative strength of the uniform ferromagnetic interaction at $\bm q=\bm0$. 
The phase diagram includes the single-$Q$ collinear (1$Q$ collinear) phase, the single-$Q$ vertical spiral (1$Q$ VS) phase, the square-lattice skyrmion crystal (SkX) phase, the single-$Q$ conical spiral (1$Q$ CS) phase, and the fully polarized ferromagnetic (FM) phase. 
The stability region of the SkX phase strongly depends on the competition between the uniform ferromagnetic interaction and the anisotropic finite-$Q$ interactions.
}
\end{center}
\end{figure}   
\unskip

We first discuss the overall structure of the low-temperature magnetic phase diagram obtained by simulated annealing. 
Figure~\ref{fig: PD} shows the phase diagram in the plane of the magnetic field $H$ and the parameter $\gamma$, which controls the relative importance of the uniform ferromagnetic interaction at $\bm q=\bm 0$. 
Five phases appear in the present parameter range: the single-$Q$ collinear (1$Q$ collinear) state, the single-$Q$ vertical spiral (1$Q$ VS) state, the square-lattice SkX state, the single-$Q$ conical spiral (1$Q$ CS) state, and the fully polarized ferromagnetic (FM) state. 
At low magnetic fields, the system first stabilizes the single-$Q$ collinear state, where the dominant modulation appears in the $z$ component. 
With increasing field, this state changes into the single-$Q$ vertical spiral state, in which the spin texture acquires transverse components while retaining the single-$Q$ character. 
In the intermediate-field region, the square-lattice SkX phase is stabilized in a finite parameter window. 
At higher fields, the SkX phase is replaced by the single-$Q$ conical spiral phase or directly by the fully polarized ferromagnetic state depending on $\gamma$.

A central feature of figure~\ref{fig: PD} is that the stability region of the SkX phase is strongly affected by the uniform ferromagnetic interaction. 
When the $\bm q=\bm 0$ interaction is relatively weak, the SkX phase occupies a wide intermediate-field window. 
{
By contrast, when the uniform ferromagnetic component is enhanced, the SkX region rapidly shrinks and eventually disappears, leading to a direct transition from a single-$Q$ modulated state to the fully polarized ferromagnetic state.
This behavior reflects the competition between the finite-$q$ anisotropic interactions, which stabilize modulated noncoplanar textures, and the tendency toward uniform spin alignment.
Since the external magnetic field also favors a uniform magnetization, it cooperates with the $\bm{q}=\bm{0}$ interaction to progressively suppress the double-$Q$ instability as $\gamma$ increases.
Consequently, the SkX phase is stabilized only when the energy gain from the double-$Q$ modulation overcomes the combined effects of the Zeeman coupling and the uniform ferromagnetic interaction.
The overall sequence of field-induced magnetic phases nevertheless remains unchanged because the finite-$q$ interactions themselves are unaffected by $\gamma$.
}

\begin{figure}[ht!]
\begin{center}
\includegraphics[width=9.0 cm]{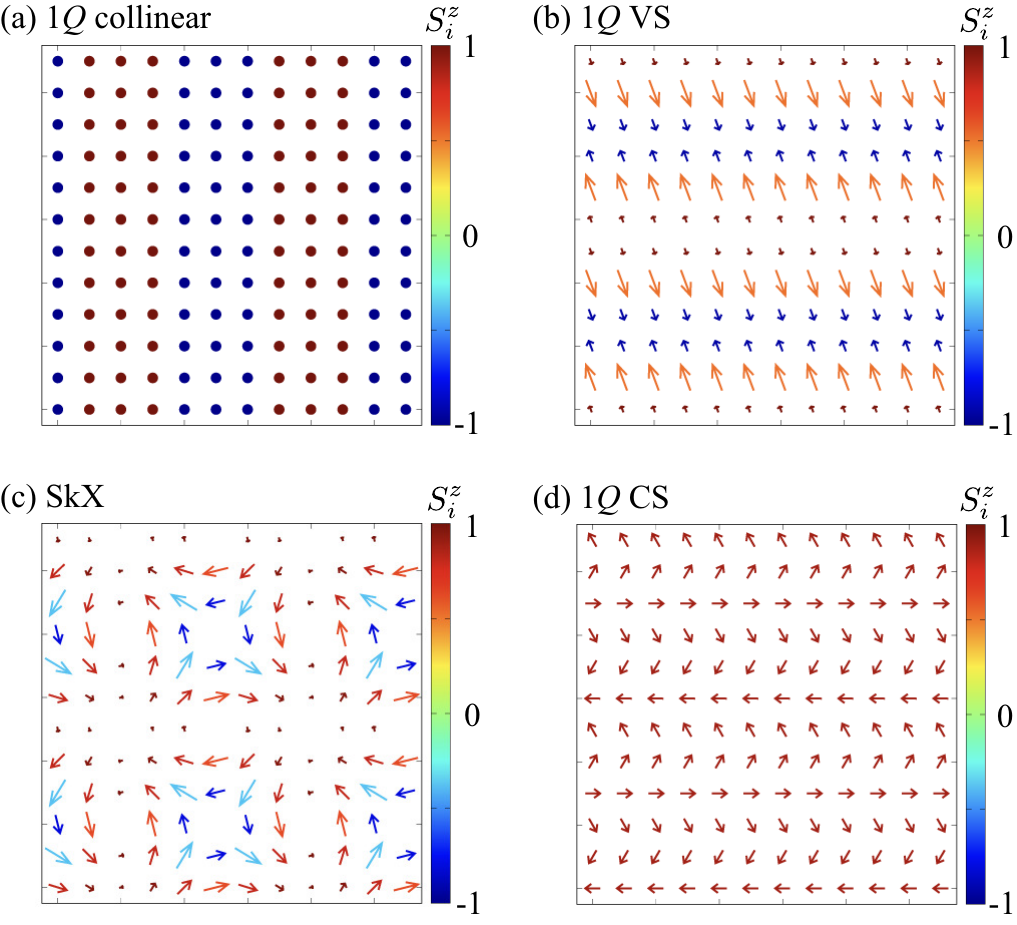}
\caption{
\label{fig: Spin}
(Colour online) Representative real-space spin configurations obtained from simulated annealing calculations. 
Panels (a)--(d) correspond to the single-$Q$ collinear (1$Q$ collinear) phase at $\gamma=0.2$ and $H=0.1$, the single-$Q$ vertical spiral (1$Q$ VS) phase at $\gamma=0.2$ and $H=0.7$, the square-lattice skyrmion crystal (SkX) phase at $\gamma=0.2$ and $H=1$, and the single-$Q$ conical spiral (1$Q$ CS) phase at $\gamma=0.2$ and $H=1.15$, respectively. 
The arrows indicate the in-plane spin components, while the color scale represents the out-of-plane component $S_i^z$. 
The SkX phase in panel (c) exhibits a characteristic double-$Q$ noncoplanar texture accompanied by periodic swirling spin structures, in contrast to the single-$Q$ modulated states in the other panels.
}
\end{center}
\end{figure}   
\unskip

Representative real-space spin configurations are shown in figure~\ref{fig: Spin}. 
Figure~\ref{fig: Spin}(a) displays the single-$Q$ collinear state. 
The spins are almost parallel or antiparallel to the field direction, and the modulation is visible only through the spatial alternation of $S_i^z$. 
Since the transverse spin components are absent, this state has no scalar spin chirality. 
Figure~\ref{fig: Spin}(b) shows the single-$Q$ vertical spiral state. 
In this phase, the spins rotate in a plane containing the field direction, forming a vertical spiral with a single modulation vector. 
Although the spin texture is noncollinear, it remains coplanar, and therefore it does not produce a net scalar spin chirality.

The SkX state shown in figure~\ref{fig: Spin}(c) is qualitatively different from these single-$Q$ states. 
The spin texture consists of a periodic array of swirling spin configurations, where the in-plane spin components wind around regions with a pronounced out-of-plane spin polarization. 
The texture is generated by the superposition of two orthogonal modulations and is accompanied by higher-harmonic Fourier components. 
The coexistence of the two fundamental modulations produces a noncoplanar spin configuration with a finite scalar spin chirality. 
{
Unlike conventional Bloch- or N\'eel-type skyrmions, which exhibit an axisymmetric spin texture with a uniform sense of in-plane rotation, the present square-lattice SkX is intrinsically non-axisymmetric owing to the superposition of two symmetry-related ordering wave vectors.
Nevertheless, the spin texture continuously wraps the unit sphere within a magnetic unit cell and carries a quantized skyrmion number $n_{\rm sk}=+1$, confirming that it is a genuine SkX.
}
It is noteworthy that the skyrmion cores are located nearly on lattice sites rather than at interstitial positions. 
This feature originates from the relatively strong easy-axis anisotropic interaction, which energetically favors a localized out-of-plane spin polarization on lattice sites~\cite{Hayami_PhysRevResearch.3.043158}. 
Figure~\ref{fig: Spin}(d) represents the single-$Q$ conical spiral state. 
In this phase, the in-plane spin components form a single-$Q$ spiral, while the $z$ component remains almost uniform due to the magnetic field. 
Thus, the single-$Q$ conical spiral state is noncoplanar in a local sense but does not possess the two-dimensional topological texture characteristic of the SkX phase.

\begin{figure}[ht!]
\begin{center}
\includegraphics[width=5.0 cm]{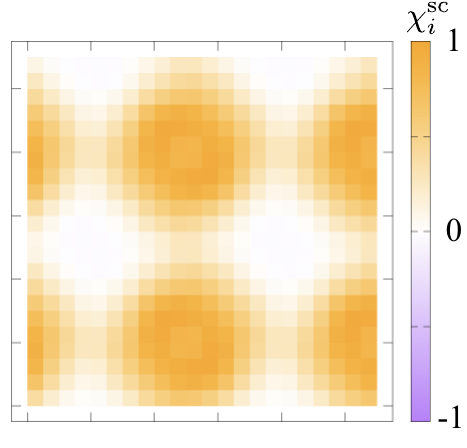}
\caption{
\label{fig: Chirality}
(Colour online) Spatial distribution of the scalar spin chirality in the square-lattice skyrmion crystal (SkX) phase obtained by simulated annealing calculations. 
The noncoplanar spin arrangement produces a periodically modulated distribution of scalar spin chirality over the magnetic unit cell.
}
\end{center}
\end{figure}   
\unskip

The topological character of the SkX phase is further confirmed by the real-space distribution of the scalar spin chirality shown in figure~\ref{fig: Chirality}. 
The scalar spin chirality exhibits a periodic pattern synchronized with the square-lattice SkX texture. 
Importantly, the scalar spin chirality distribution has a uniform sign over the magnetic unit cell, leading to a nonzero spatial average. 
This behavior distinguishes the SkX phase from other noncollinear states in the phase diagram, such as the single-$Q$ vertical spiral and single-$Q$ conical spiral states. 
In the former state, the scalar spin chirality is absent both locally and globally, whereas in the latter state, finite local scalar spin chirality develops but its spatial average remains zero.
The finite scalar spin chirality in figure~\ref{fig: Chirality} reflects the solid angle subtended by the neighboring spins and is the microscopic origin of the topological nature of the SkX state.

\begin{figure}[ht!]
\begin{center}
\includegraphics[width=14.0 cm]{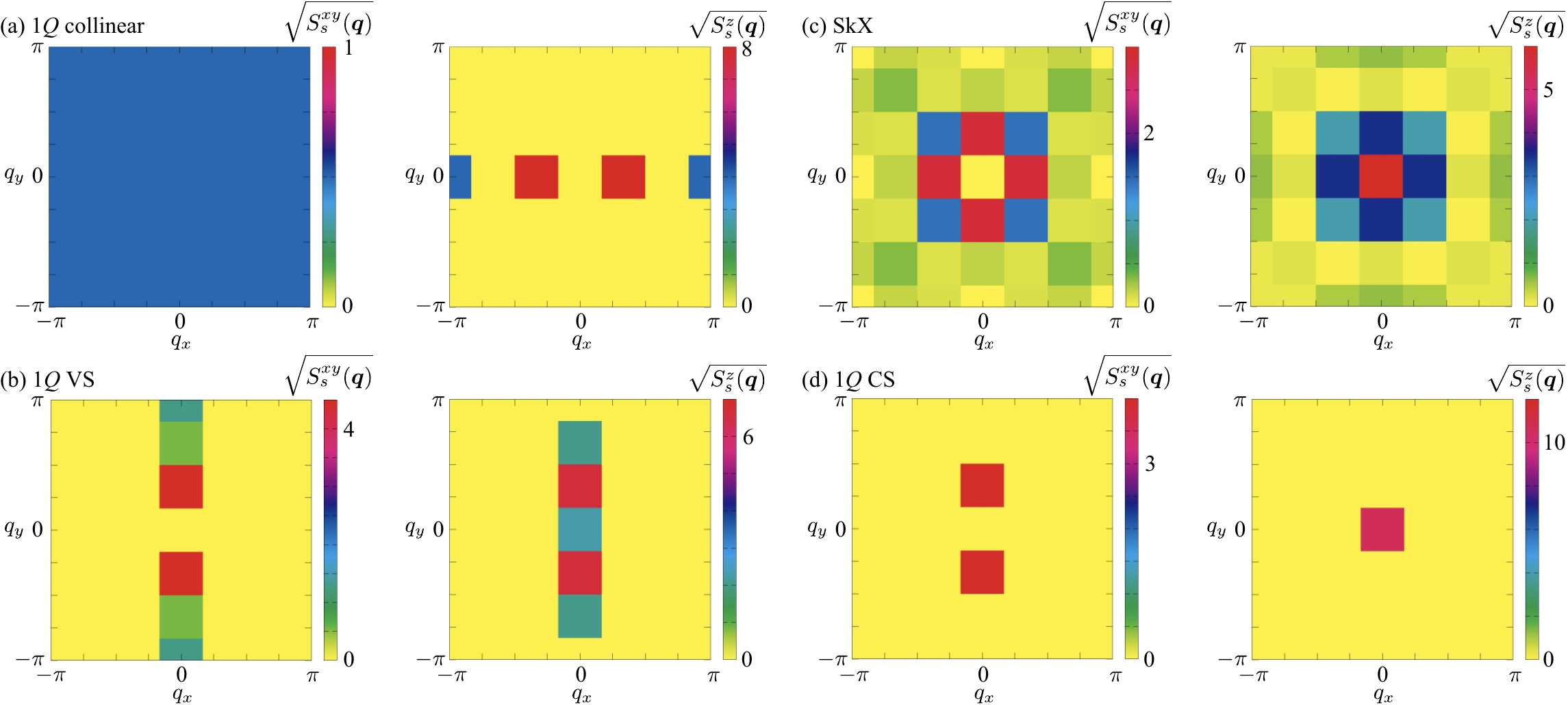}
\caption{
\label{fig: Sq}
(Colour online) Spin structure factors for representative magnetic phases obtained in the present model. 
Panels (a)--(d) correspond to the single-$Q$ collinear (1$Q$ collinear) phase, the single-$Q$ vertical spiral (1$Q$ VS) phase, the square-lattice skyrmion crystal (SkX) phase, and the single-$Q$ conical spiral (1$Q$ CS) phase, respectively. 
The parameters are the same as those in figure~\ref{fig: Spin}. 
The intensity plots display the momentum-space distribution of the magnetic modulations. 
The single-$Q$ phases exhibit the dominant spectral weight at a single ordering wave vector, whereas the SkX phase is characterized by simultaneous intensities at two orthogonal wave vectors together with finite higher-harmonic wave-vector components originating from the double-$Q$ superposition.
}
\end{center}
\end{figure}   
\unskip

The corresponding momentum-space information is summarized in figure~\ref{fig: Sq}. 
In figure~\ref{fig: Sq}(a), the single-$Q$ collinear state shows Bragg peaks only in the $z$ component at the fundamental ordering wave vector. 
The absence of the in-plane spectral weight is consistent with the real-space collinear texture in figure~\ref{fig: Spin}(a). 
In figure~\ref{fig: Sq}(b), the single-$Q$ vertical spiral state exhibits both in-plane and out-of-plane peaks at a single ordering wave-vector direction, reflecting the coplanar spiral modulation. 
Thus, the transition from the single-$Q$ collinear state to the single-$Q$ vertical spiral state can be understood as the development of transverse spin components while keeping the single-$Q$ character.

Figure~\ref{fig: Sq}(c) shows the structure factor in the SkX phase. 
In contrast to the single-$Q$ states, the spectral weight appears at both orthogonal ordering wave vectors. 
In addition, finite intensities are found at higher-harmonic wave vectors related to $\bm Q_1\pm \bm Q_2$. 
This momentum-space structure is a clear signature of the double-$Q$ nature of the square-lattice SkX state. 
The higher-harmonic wave-vector components are not merely secondary features; they help shape the square-symmetric spin texture and reinforce the stability of the noncoplanar double-$Q$ state. 
Figure~\ref{fig: Sq}(d) displays the single-$Q$ conical spiral state, where the in-plane component retains a single-$Q$ peak, while the out-of-plane component is dominated by the uniform component associated with the field-induced magnetization. 
This spectral feature is consistent with the conical spiral texture in figure~\ref{fig: Spin}(d).

\begin{figure}[ht!]
\begin{center}
\includegraphics[width=8.0 cm]{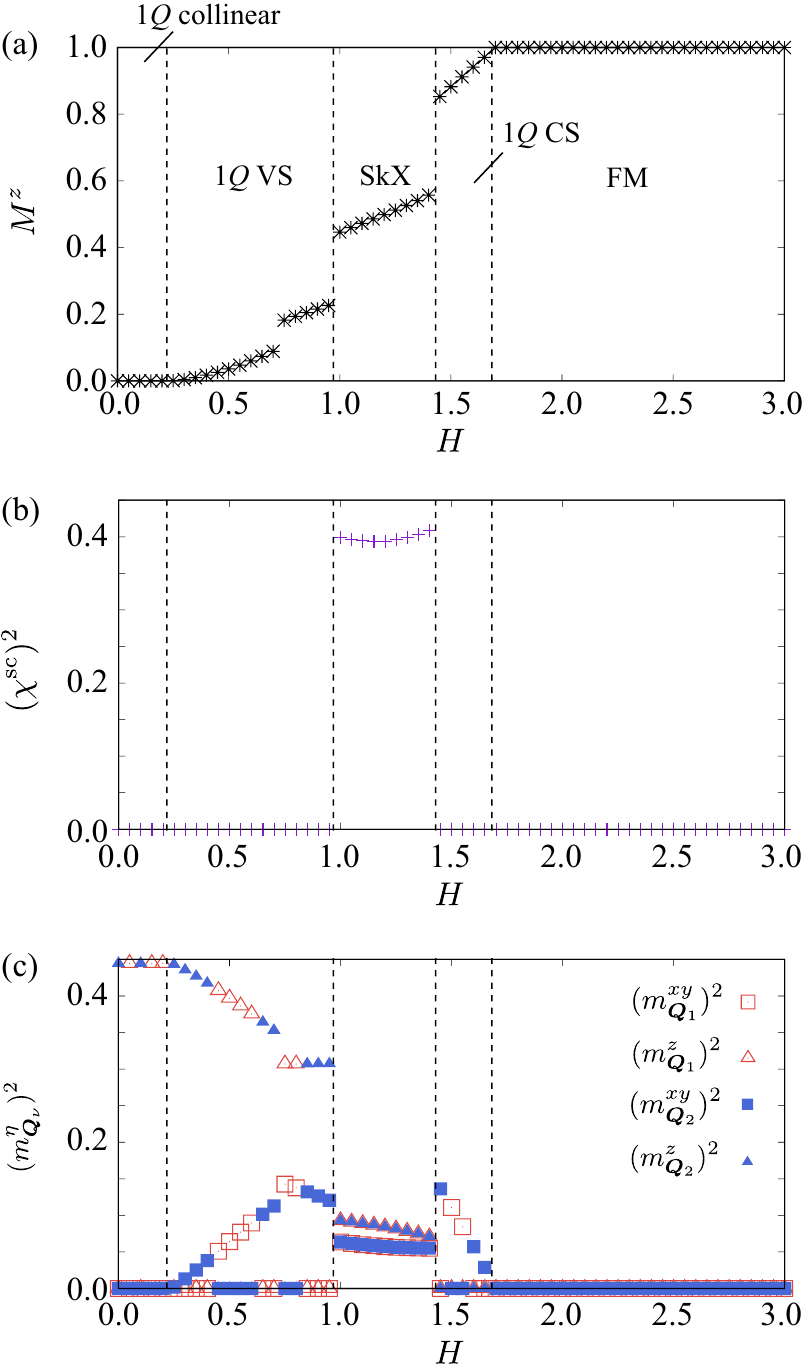}
\caption{
\label{fig: mag_gamma=0.0}
(Colour online) Magnetic-field dependence of the physical quantities for a parameter set at $\gamma=0$ where all modulated phases appear successively before entering the fully polarized ferromagnetic state. 
(a) Magnetization $M^z$, (b) squared scalar spin chirality $(\chi^{\mathrm{sc}})^2$, and (c) momentum-resolved magnetic amplitudes at the relevant ordering wave vectors $(m^\eta_{\bm{Q}_\nu})^2$. 
The vertical dashed lines indicate the phase boundaries determined from anomalies in these quantities. 
The finite scalar spin chirality appearing only in the intermediate-field region identifies the square-lattice skyrmion crystal (SkX) phase, while the changes in the Fourier amplitudes reflect the reconstruction between single-$Q$ and double-$Q$ magnetic states.
}
\end{center}
\end{figure}   
\unskip

We next examine the field evolution of the magnetic observables. 
Figure~\ref{fig: mag_gamma=0.0} presents the representative case where all four modulated phases appear sequentially before the system reaches the fully polarized ferromagnetic state. 
The magnetization in figure~\ref{fig: mag_gamma=0.0}(a) is almost zero in the single-$Q$ collinear state, starts to increase in the single-$Q$ vertical spiral state, and exhibits a clear jump at the transition to the SkX phase. 
The discontinuity in $M^z$ gradually indicates that the single-$Q$ vertical spiral--SkX transition is first order. 
Inside the SkX phase, $M^z$ increases  with the magnetic field, reflecting the progressive enlargement of the field-polarized spin component. 
The transition from the SkX phase to the single-$Q$ conical spiral phase is accompanied by another abrupt change in magnetization, suggesting a first-order character. 
At still higher fields, the magnetization approaches saturation as the system enters the fully polarized ferromagnetic phase.

The scalar spin chirality in figure~\ref{fig: mag_gamma=0.0}(b) provides a direct diagnostic of the SkX phase. 
It is zero in the single-$Q$ collinear and single-$Q$ vertical spiral phases, becomes finite only in the SkX phase, and disappears again in the single-$Q$ conical spiral and fully polarized ferromagnetic phases. 
The onset and disappearance of $\chi^{\rm sc}$ occur discontinuously at the phase boundaries, again supporting the first-order nature of the transitions involving the SkX state. 
Figure~\ref{fig: mag_gamma=0.0}(c) shows the Fourier amplitudes at $\bm Q_1$ and $\bm Q_2$. 
In the single-$Q$ collinear and single-$Q$ vertical spiral phases, one ordering wave vector dominates, whereas the other remains almost absent. 
In the SkX phase, both $\bm Q_1$ and $\bm Q_2$ components become finite, demonstrating the double-$Q$ character. 
In the single-$Q$ conical spiral phase, the double-$Q$ balance is lost, and the system returns to a single-$Q$ modulation before becoming fully polarized.

\begin{figure}[ht!]
\begin{center}
\includegraphics[width=8.0 cm]{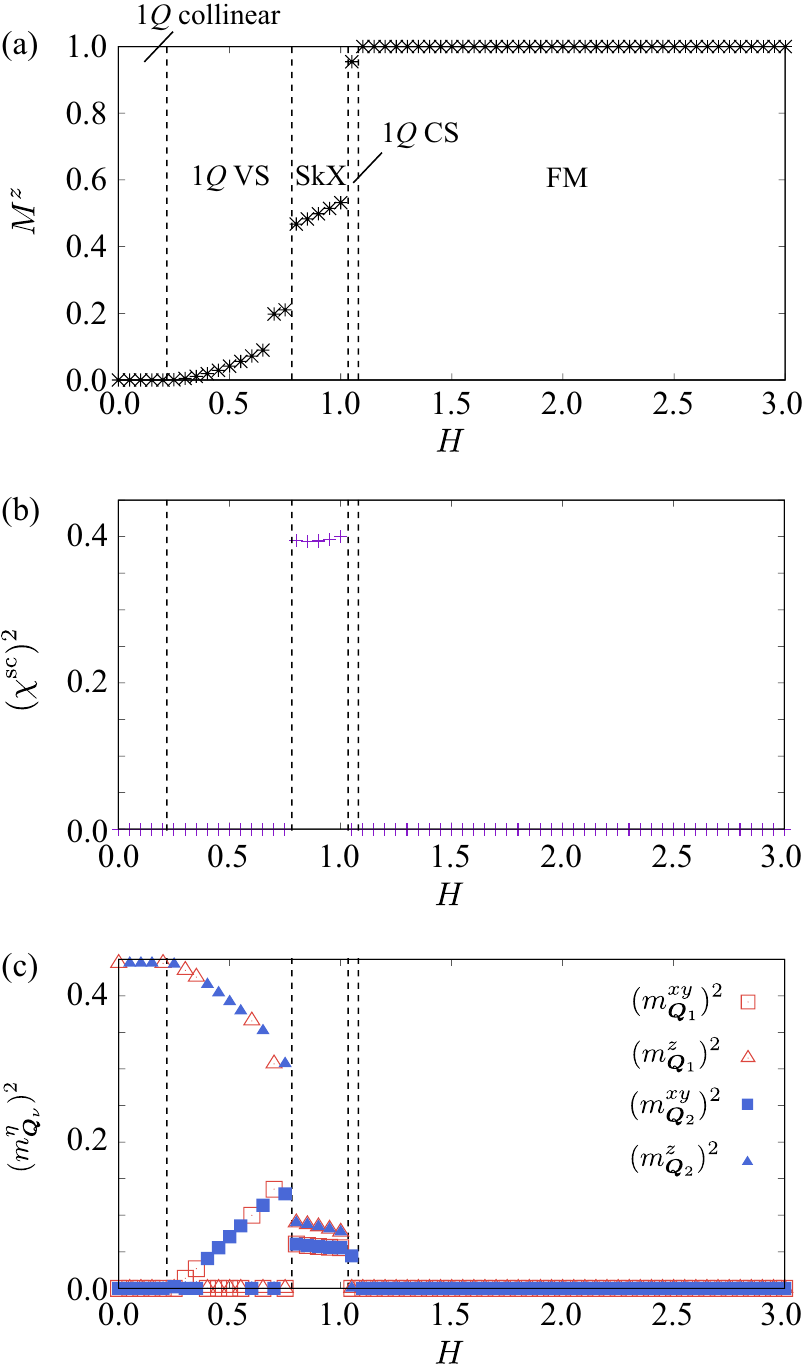}
\caption{
\label{fig: mag_gamma=0.3}
(Colour online) Magnetic-field dependence of (a) $M^z$, (b) $(\chi^{\mathrm{sc}})^2$, and (c) $(m^\eta_{\bm{Q}_\nu})^2$ at $\gamma=0.3$. 
Compared with figure~\ref{fig: mag_gamma=0.0}, the stability region of the square-lattice skyrmion crystal (SkX) phase is reduced due to the enhanced uniform ferromagnetic interaction.
}
\end{center}
\end{figure}   
\unskip

Figure~\ref{fig: mag_gamma=0.3} shows another representative field scan at $\gamma=0.3$.
{
Compared with the $\gamma=0$ case in figure~\ref{fig: mag_gamma=0.0}, the overall sequence of magnetic phases remains unchanged, whereas the magnetic-field window of the SkX phase is significantly reduced.
As shown in figure~\ref{fig: mag_gamma=0.3}(a), the magnetization exhibits sharper changes around the SkX phase, reflecting the enhanced competition between the uniform ferromagnetic interaction and the double-$Q$ instability.
Correspondingly, the scalar spin chirality in figure~\ref{fig: mag_gamma=0.3}(b) remains finite only over a narrower field range, indicating that the topological phase is destabilized as the balance shifts toward uniform magnetization.
The Fourier amplitudes in figure~\ref{fig: mag_gamma=0.3}(c) further show that the coexistence region of the two ordering wave vectors, $\bm{Q}_1$ and $\bm{Q}_2$, is substantially reduced, while the adjacent phases remain dominated by a single modulation.
These results demonstrate that increasing $\gamma$, namely strengthening the $\bm{q}=\bm{0}$ ferromagnetic interaction, primarily suppresses the double-$Q$ instability without qualitatively changing the sequence of field-induced phase transitions.
}

\begin{figure}[ht!]
\begin{center}
\includegraphics[width=9.0 cm]{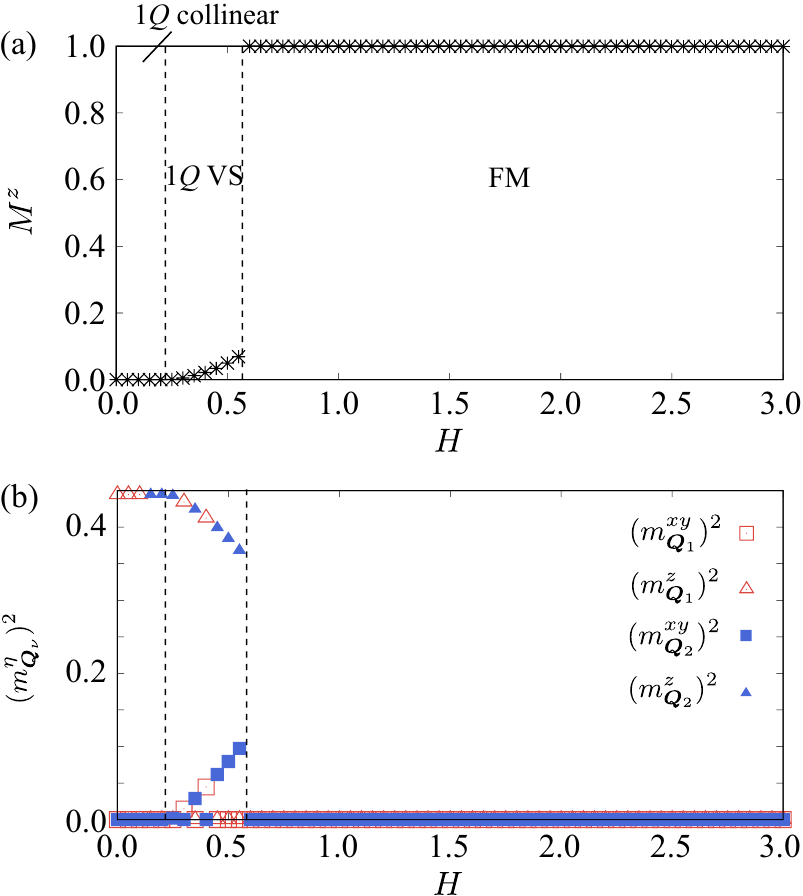}
\caption{
\label{fig: mag_gamma=0.6}
(Colour online) Magnetic-field dependence of (a) $M^z$ and (b) $(m^\eta_{\bm{Q}_\nu})^2$ at $\gamma=0.6$. 
The square-lattice skyrmion crystal (SkX) phase disappears in this parameter region, and the system undergoes a direct transition from the single-$Q$ vertical spiral state to the fully polarized ferromagnetic state.
}
\end{center}
\end{figure}   
\unskip

For a stronger uniform ferromagnetic tendency at $\gamma=0.6$, the SkX phase disappears, as shown in figure~\ref{fig: mag_gamma=0.6}. 
The system first enters the single-$Q$ collinear state at low fields and then changes into the single-$Q$ vertical spiral state. 
However, instead of forming the SkX phase at intermediate fields, the system undergoes a direct transition into the fully polarized ferromagnetic state. 
The magnetization in figure~\ref{fig: mag_gamma=0.6}(a) exhibits a sudden jump to the saturated value, which indicates a strong first-order transition. 
Since the scalar spin chirality is absent in the entire field range, no topological double-$Q$ phase is stabilized. 
The Fourier amplitudes in figure~\ref{fig: mag_gamma=0.6}(b) also show that the spectral weight remains single-$Q$ before it collapses at the transition to the fully polarized ferromagnetic state. 
This result demonstrates that a sufficiently strong $\bm q=\bm0$ interaction suppresses the noncoplanar double-$Q$ instability and favors a direct single-$Q$--ferromagnetic transition.

\begin{figure}[ht!]
\begin{center}
\includegraphics[width=9.0 cm]{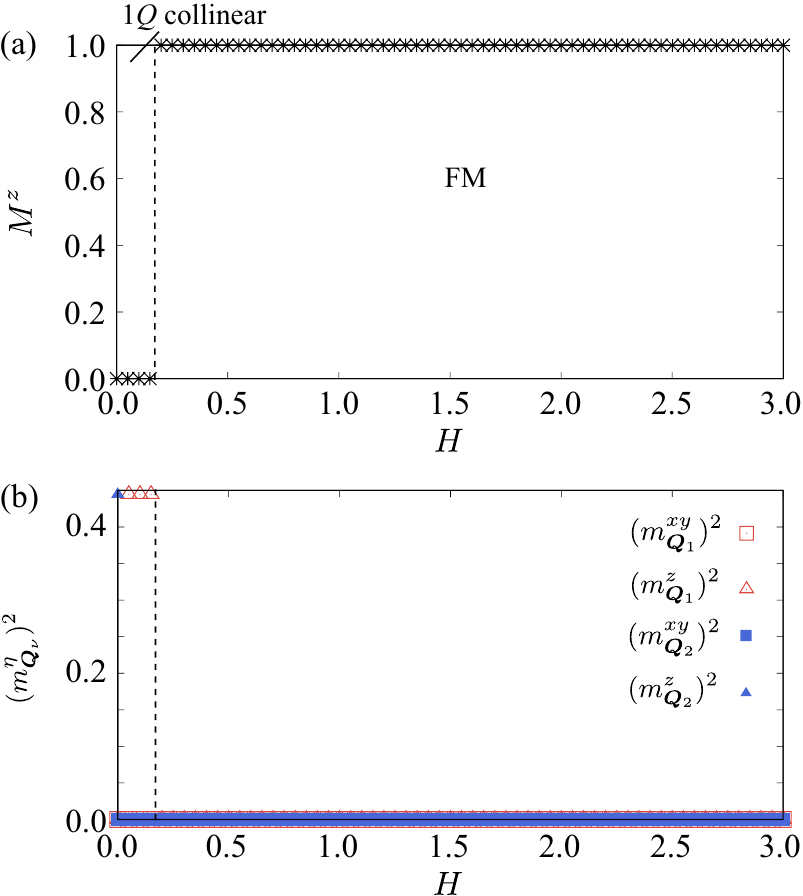}
\caption{
\label{fig: mag_gamma=1.0}
(Colour online) Magnetic-field dependence of (a) $M^z$ and (b) $(m^\eta_{\bm{Q}_\nu})^2$ at $\gamma=1$. 
For strong uniform ferromagnetic interaction, the system exhibits a direct first-order transition from the single-$Q$ collinear phase to the fully polarized ferromagnetic state without stabilizing intermediate noncoplanar phases.
}
\end{center}
\end{figure}   
\unskip

Figure~\ref{fig: mag_gamma=1.0} represents the limiting case where the uniform ferromagnetic interaction at $\gamma=1$ is so strong that even the single-$Q$ vertical spiral region is almost completely eliminated. 
As shown in figure~\ref{fig: mag_gamma=1.0}(a), the magnetization jumps directly from the single-$Q$ collinear state to the fully polarized ferromagnetic state at a low magnetic field. 
The transition is strongly first order, reflecting the direct competition between the finite-$Q$ collinear modulation and the uniform ferromagnetic state. 
The Fourier amplitudes in figure~\ref{fig: mag_gamma=1.0}(b) vanish abruptly at the same transition field. 
This behavior indicates that the finite-$Q$ modulation cannot survive once the Zeeman energy and the uniform ferromagnetic interaction cooperate to polarize the spins. 
In this regime, the anisotropic finite-$Q$ interaction is insufficient to stabilize either a vertical spiral, a conical spiral, or the SkX phase.

\begin{figure}[ht!]
\begin{center}
\includegraphics[width=9.0 cm]{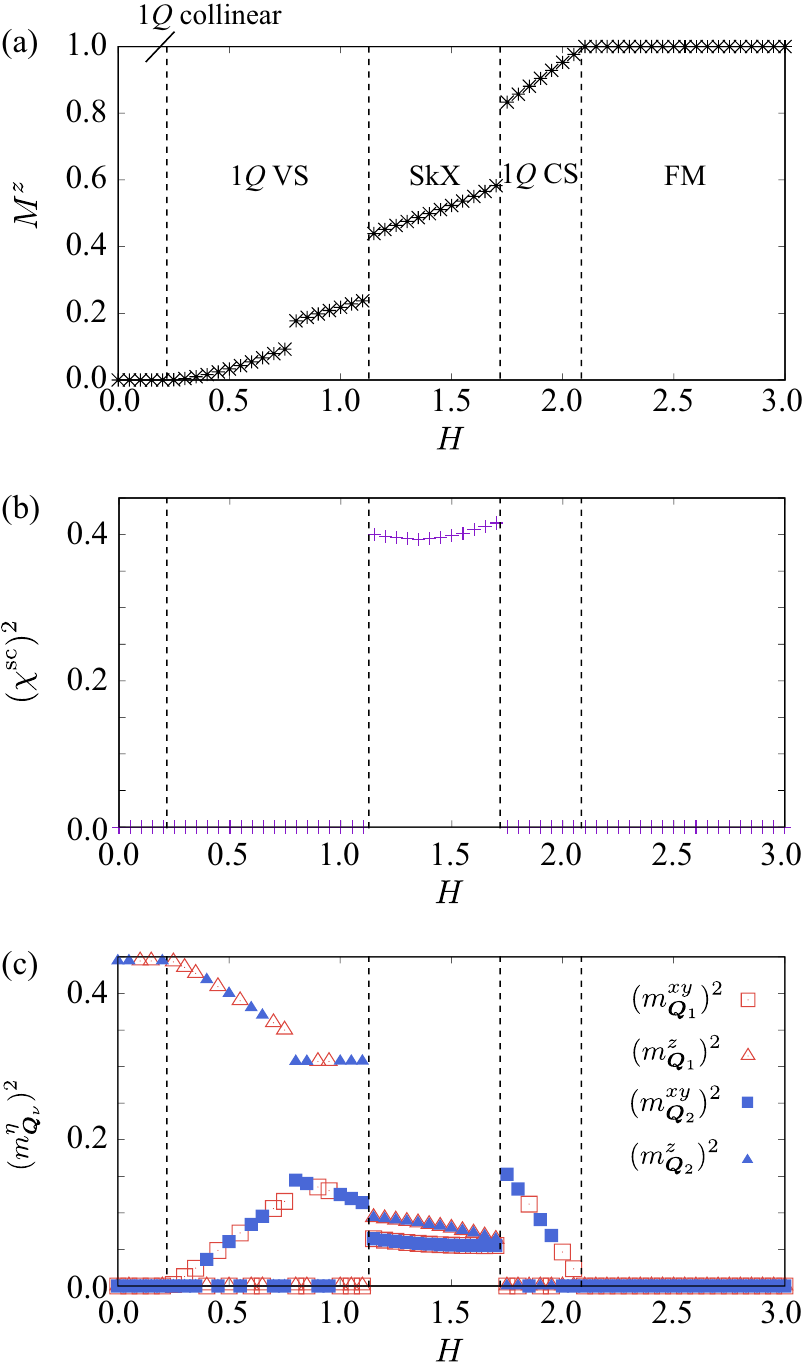}
\caption{
\label{fig: mag_gamma=-0.2}
(Colour online) Magnetic-field dependence of (a) $M^z$, (b) $(\chi^{\mathrm{sc}})^2$, and (c) $(m^\eta_{\bm{Q}_\nu})^2$ at $\gamma=-0.2$. 
Compared with the cases for positive $\gamma$, the square-lattice skyrmion crystal (SkX) phase extends over a wider magnetic-field region, and the single-$Q$ conical spiral phase is stabilized in the high-field regime.
}
\end{center}
\end{figure}   
\unskip

\begin{figure}[ht!]
\begin{center}
\includegraphics[width=9.0 cm]{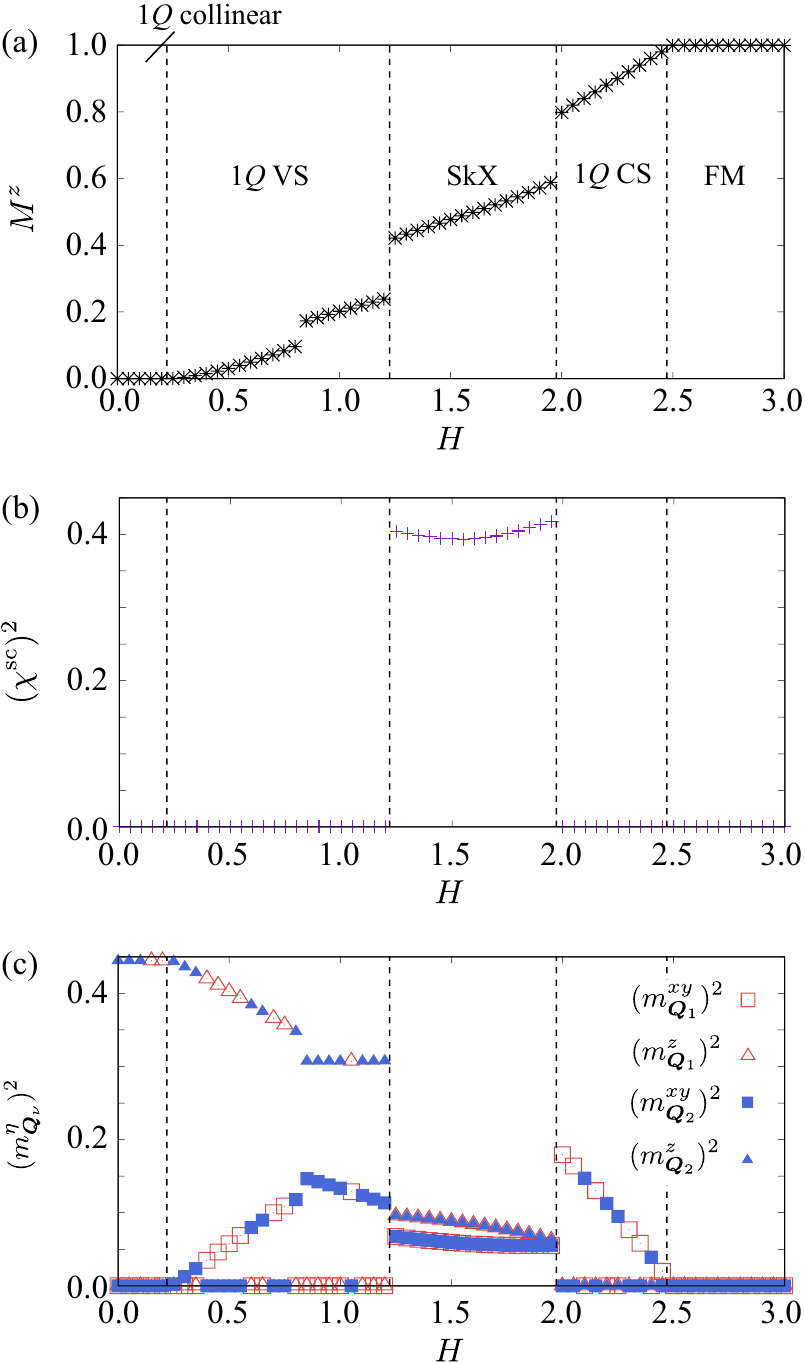}
\caption{
\label{fig: mag_gamma=-0.4}
(Colour online) Magnetic-field dependence of (a) $M^z$, (b) $(\chi^{\mathrm{sc}})^2$, and (c) $(m^\eta_{\bm{Q}_\nu})^2$ at $\gamma=-0.4$. 
The reduction of the uniform ferromagnetic interaction further enlarges the stability region of the square-lattice skyrmion crystal (SkX) phase and broadens the field range of the single-$Q$ conical spiral phase before entering the fully polarized ferromagnetic state.
}
\end{center}
\end{figure}

We finally discuss the opposite regime, where the uniform ferromagnetic interaction is weakened. 
Figures~\ref{fig: mag_gamma=-0.2} and \ref{fig: mag_gamma=-0.4} show that the SkX phase becomes more robust and that the high-field single-$Q$ conical spiral phase is also stabilized over a wider field interval. 
In figure~\ref{fig: mag_gamma=-0.2}(a) for $\gamma=-0.2$, the magnetization increases through the sequence single-$Q$ collinear, single-$Q$ vertical spiral, SkX, single-$Q$ conical spiral, and fully polarized ferromagnetic states. 
Compared with figures~\ref{fig: mag_gamma=0.0} and \ref{fig: mag_gamma=0.3}, the SkX phase extends to higher fields. 
The finite scalar spin chirality in figure~\ref{fig: mag_gamma=-0.2}(b) appears over a broader field window, demonstrating the enhanced stability of the topological double-$Q$ state. 
The Fourier amplitudes in figure~\ref{fig: mag_gamma=-0.2}(c) show sizable contributions from both $\bm Q_1$ and $\bm Q_2$ in the SkX phase, whereas the single-$Q$ conical spiral phase is characterized by the recovery of a single dominant modulation together with a large uniform magnetization.

\unskip

This tendency becomes even clearer in figure~\ref{fig: mag_gamma=-0.4} for $\gamma=-0.4$. 
The SkX phase occupies a broad intermediate-field region, and the single-$Q$ conical spiral phase persists up to higher fields before the system becomes fully polarized. 
The magnetization in figure~\ref{fig: mag_gamma=-0.4}(a) changes stepwise at the phase boundaries, while the scalar spin chirality in figure~\ref{fig: mag_gamma=-0.4}(b) remains finite only in the SkX phase. 
The Fourier amplitudes in figure~\ref{fig: mag_gamma=-0.4}(c) reveal that the double-$Q$ spectral weight is well developed in the SkX phase. 
After the transition to the single-$Q$ conical spiral phase, one of the modulations becomes dominant, and the system gradually approaches the fully polarized ferromagnetic state with increasing field. 
These results indicate that the weakening of the $\bm q=\bm0$ ferromagnetic interaction allows the finite-$Q$ and higher-harmonic wave-vector interactions to control the magnetic texture over a wider field range.

The above results establish the following physical picture. 
The anisotropic finite-$Q$ interaction favors modulated states with a sizable $z$-spin component, while the higher-harmonic wave-vector interaction supports the coexistence of orthogonal modulations. 
Their cooperation stabilizes the square-lattice SkX phase. 
The uniform $\bm q=\bm0$ ferromagnetic interaction, on the other hand, competes with this mechanism by favoring collinear spin alignment. 
When this uniform component is too strong, the double-$Q$ SkX phase is suppressed, and the system undergoes a first-order transition from a single-$Q$ state to the fully polarized ferromagnetic state. 
When it is weakened, the double-$Q$ instability becomes more effective, leading to an enlarged SkX region and the appearance of the single-$Q$ conical spiral phase at high fields. 
Thus, the stability of square-lattice SkX states in centrosymmetric magnets is controlled by a delicate balance among uniform ferromagnetism, anisotropic finite-wave-vector instability, and higher-harmonic wave-vector interactions.

\section{Conclusions}
\label{sec: Conclusions}

In summary, we have investigated the stability of square-lattice SkX states in a centrosymmetric spin model with momentum-dependent magnetic interactions. 
The model contains four essential ingredients: the uniform ferromagnetic interaction at $\bm q=\bm0$, the easy-axis anisotropic interactions at the fundamental wave vectors $\bm Q_1$ and $\bm Q_2$, the higher-harmonic wave-vector interactions at $\bm Q_1+\bm Q_2$ and $\bm Q_1-\bm Q_2$, and the Zeeman coupling to an external magnetic field. 
By performing simulated annealing calculations, we constructed the low-temperature magnetic phase diagram as a function of the magnetic field and the parameter $\gamma$, which controls the strength of the uniform ferromagnetic interaction.

We found that the square-lattice SkX phase is stabilized in the intermediate-field region when the uniform ferromagnetic interaction is sufficiently weak. 
The SkX state is characterized by a double-$Q$ noncoplanar spin texture with finite scalar spin chirality and simultaneous spectral weights at two orthogonal ordering wave vectors together with higher-harmonic wave-vector components. 
A central result of the present study is that the uniform $\bm q=\bm0$ ferromagnetic interaction strongly controls the stability of the SkX phase. 
For larger positive $\gamma$, the uniform component suppresses the double-$Q$ texture, causing the SkX phase to shrink or disappear through direct first-order transitions from the single-$Q$ modulated phases to the fully polarized ferromagnetic state. 
On the other hand, reducing the uniform ferromagnetic interaction enlarges the SkX region and stabilizes the single-$Q$ conical spiral phase in the high-field regime.

These results clarify that the square-lattice SkX phase in centrosymmetric systems is governed by a delicate balance among uniform ferromagnetism, easy-axis finite-wave-vector instability, and higher-harmonic wave-vector interactions. 
The uniform ferromagnetic interaction acts as a competing factor against the double-$Q$ SkX formation, whereas the finite-$Q$ anisotropic and higher-harmonic wave-vector interactions cooperate to stabilize the noncoplanar topological texture. 
The present findings, therefore, provide a useful guideline for understanding and controlling square-lattice SkX states in centrosymmetric magnets, especially in itinerant or frustrated systems where effective magnetic interactions are naturally structured in momentum space.

\section*{Acknowledgment}
\sloppy
This research was supported by JSPS KAKENHI Grant Numbers JP22H00101, JP22H01183, JP23H04869, JP23K03288, JP23K20827, and by JST CREST (JPMJCR23O4) and JST FOREST (JPMJFR2366).


\bibliographystyle{cmpj}
\bibliography{ref}

\begin{thebibliography}{100}
\providecommand{\url}[1]{\texttt{#1}}
\providecommand{\urlprefix}{URL }
\expandafter\ifx\csname urlstyle\endcsname\relax
  \providecommand{\doi}[1]{doi:\discretionary{}{}{}#1}\else
  \providecommand{\doi}{doi:\discretionary{}{}{}\begingroup
  \urlstyle{rm}\Url}\fi
\providecommand{\eprint}[2][]{\url{#2}}

\bibitem{nagaosa2013topological}
Nagaosa~N., Tokura~Y., Nat. Nanotechnol., 2013, \textbf{8}, No.~12, 899--911,
  \doi{10.1038/nnano.2013.243}.

\bibitem{Tokura_doi:10.1021/acs.chemrev.0c00297}
Tokura~Y., Kanazawa~N., Chem. Rev., 2021, \textbf{121}, 2857,
  \doi{10.1021/acs.chemrev.0c00297}.

\bibitem{gobel2021beyond}
G{\"o}bel~B., Mertig~I., Tretiakov~O.~A., Phys. Rep., 2021, \textbf{895}, 1,
  \doi{10.1016/j.physrep.2020.10.001}.

\bibitem{guslienko2024magnetic}
Guslienko~K., Magnetism, 2024, \textbf{4}, No.~4, 383--399,
  \doi{10.3390/magnetism4040025}.

\bibitem{Jonietz_skyrmion}
Jonietz~F., M$\ddot{\rm u}$hlbauer~S., Pfleiderer~C., Neubauer~A., M$\ddot{\rm
  u}$nzer~W., Bauer~A., Adams~T., Georgii~R., B$\ddot{\rm o}$ni~P.,
  Duine~R.~A., Everschor~K., Garst~M., Rosch~A., Science, 2010, \textbf{330},
  1648, \doi{10.1126/science.1195709}.

\bibitem{yu2012skyrmion}
Yu~X.~Z., Kanazawa~N., Zhang~W., Nagai~T., Hara~T., Kimoto~K., Matsui~Y.,
  Onose~Y., Tokura~Y., Nat. Commun., 2012, \textbf{3}, 988,
  \doi{10.1038/ncomms1990}.

\bibitem{fert2013skyrmions}
Fert~A., Cros~V., Sampaio~J., Nat. Nanotechnol., 2013, \textbf{8}, No.~3, 152,
  \doi{10.1038/nnano.2013.29}.

\bibitem{zhang2015magnetic}
Zhang~X., Ezawa~M., Zhou~Y., Sci. Rep., 2015, \textbf{5}, 9400,
  \doi{10.1038/srep09400}.

\bibitem{fert2017magnetic}
Fert~A., Reyren~N., Cros~V., Nat. Rev. Mater., 2017, \textbf{2}, No.~7, 17031,
  \doi{10.1038/natrevmats.2017.31}.

\bibitem{jiang2017direct}
Jiang~W., Zhang~X., Yu~G., Zhang~W., Wang~X., Benjamin~Jungfleisch~M.,
  Pearson~J.~E., Cheng~X., Heinonen~O., Wang~K.~L., Zhou~Y., Hoffmann~A., {te
  Velthuis}~S. G.~E., Nat. Phys., 2017, \textbf{13}, No.~2, 162--169,
  \doi{10.1038/nphys3883}.

\bibitem{Baltz_RevModPhys.90.015005}
Baltz~V., Manchon~A., Tsoi~M., Moriyama~T., Ono~T., Tserkovnyak~Y., Rev. Mod.
  Phys., 2018, \textbf{90}, 015005, \doi{10.1103/RevModPhys.90.015005}.

\bibitem{luo2018reconfigurable}
Luo~S., Song~M., Li~X., Zhang~Y., Hong~J., Yang~X., Zou~X., Xu~N., You~L., Nano
  Lett., 2018, \textbf{18}, No.~2, 1180--1184,
  \doi{10.1021/acs.nanolett.7b04722}.

\bibitem{Chauwin_PhysRevApplied.12.064053}
Chauwin~M., Hu~X., Garcia-Sanchez~F., Betrabet~N., Paler~A., Moutafis~C.,
  Friedman~J.~S., Phys. Rev. Appl., 2019, \textbf{12}, 064053,
  \doi{10.1103/PhysRevApplied.12.064053}.

\bibitem{zhang2020skyrmion}
Zhang~X., Zhou~Y., Song~K.~M., Park~T.-E., Xia~J., Ezawa~M., Liu~X., Zhao~W.,
  Zhao~G., Woo~S., J. Phys.: Condens. Matter, 2020, \textbf{32}, No.~14,
  143001, \doi{10.1088/1361-648X/ab5488}.

\bibitem{yu2020motion}
Yu~X., Morikawa~D., Nakajima~K., Shibata~K., Kanazawa~N., Arima~T.-h.,
  Nagaosa~N., Tokura~Y., Sci. Adv., 2020, \textbf{6}, No.~25, eaaz9744,
  \doi{10.1126/sciadv.aaz9744}.

\bibitem{dzyaloshinsky1958thermodynamic}
Dzyaloshinsky~I., J. Phys. Chem. Solids, 1958, \textbf{4}, No.~4, 241--255,
  \doi{10.1016/0022-3697(58)90076-3}.

\bibitem{moriya1960anisotropic}
Moriya~T., Phys. Rev., 1960, \textbf{120}, No.~1, 91,
  \doi{10.1103/PhysRev.120.91}.

\bibitem{rossler2006spontaneous}
R{\"o}{\ss}ler~U.~K., Bogdanov~A.~N., Pfleiderer~C., Nature, 2006,
  \textbf{442}, No. 7104, 797--801, \doi{10.1038/nature05056}.

\bibitem{Binz_PhysRevLett.96.207202}
Binz~B., Vishwanath~A., Aji~V., Phys. Rev. Lett., 2006, \textbf{96}, 207202,
  \doi{10.1103/PhysRevLett.96.207202}.

\bibitem{Binz_PhysRevB.74.214408}
Binz~B., Vishwanath~A., Phys. Rev. B, 2006, \textbf{74}, 214408,
  \doi{10.1103/PhysRevB.74.214408}.

\bibitem{Yi_PhysRevB.80.054416}
Yi~S.~D., Onoda~S., Nagaosa~N., Han~J.~H., Phys. Rev. B, 2009, \textbf{80},
  054416, \doi{10.1103/PhysRevB.80.054416}.

\bibitem{Butenko_PhysRevB.82.052403}
Butenko~A.~B., Leonov~A.~A., R\"o\ss{}ler~U.~K., Bogdanov~A.~N., Phys. Rev. B,
  2010, \textbf{82}, 052403, \doi{10.1103/PhysRevB.82.052403}.

\bibitem{heinze2011spontaneous}
Heinze~S., von Bergmann~K., Menzel~M., Brede~J., Kubetzka~A., Wiesendanger~R.,
  Bihlmayer~G., Bl{\"u}gel~S., Nat. Phys., 2011, \textbf{7}, No.~9, 713--718,
  \doi{10.1038/nphys2045}.

\bibitem{Wilson_PhysRevB.89.094411}
Wilson~M.~N., Butenko~A.~B., Bogdanov~A.~N., Monchesky~T.~L., Phys. Rev. B,
  2014, \textbf{89}, 094411, \doi{10.1103/PhysRevB.89.094411}.

\bibitem{Muhlbauer_2009skyrmion}
M{\"u}hlbauer~S., Binz~B., Jonietz~F., Pfleiderer~C., Rosch~A., Neubauer~A.,
  Georgii~R., B{\"o}ni~P., Science, 2009, \textbf{323}, No. 5916, 915--919,
  \doi{10.1126/science.1166767}.

\bibitem{Neubauer_PhysRevLett.102.186602}
Neubauer~A., Pfleiderer~C., Binz~B., Rosch~A., Ritz~R., Niklowitz~P.~G.,
  B\"oni~P., Phys. Rev. Lett., 2009, \textbf{102}, 186602,
  \doi{10.1103/PhysRevLett.102.186602}.

\bibitem{Adams_PhysRevLett.107.217206}
Adams~T., M\"uhlbauer~S., Pfleiderer~C., Jonietz~F., Bauer~A., Neubauer~A.,
  Georgii~R., B\"oni~P., Keiderling~U., Everschor~K., Garst~M., Rosch~A., Phys.
  Rev. Lett., 2011, \textbf{107}, 217206, \doi{10.1103/PhysRevLett.107.217206}.

\bibitem{Morikawa_PhysRevB.88.024408}
Morikawa~D., Shibata~K., Kanazawa~N., Yu~X.~Z., Tokura~Y., Phys. Rev. B, 2013,
  \textbf{88}, 024408, \doi{10.1103/PhysRevB.88.024408}.

\bibitem{Bauer_PhysRevLett.110.177207}
Bauer~A., Garst~M., Pfleiderer~C., Phys. Rev. Lett., 2013, \textbf{110},
  177207, \doi{10.1103/PhysRevLett.110.177207}.

\bibitem{yu2010real}
Yu~X.~Z., Onose~Y., Kanazawa~N., Park~J.~H., Han~J.~H., Matsui~Y., Nagaosa~N.,
  Tokura~Y., Nature, 2010, \textbf{465}, No. 7300, 901--904,
  \doi{10.1038/nature09124}.

\bibitem{adams2010skyrmion}
Adams~T., M{\"u}hlbauer~S., Neubauer~A., M{\"u}nzer~W., Jonietz~F., Georgii~R.,
  Pedersen~B., B{\"o}ni~P., Rosch~A., Pfleiderer~C., J. Phys. Conf. Ser., 2010,
  \textbf{200}, No.~3, 032001, \doi{10.1088/1742-6596/200/3/032001}.

\bibitem{Munzer_PhysRevB.81.041203}
M\"unzer~W., Neubauer~A., Adams~T., M\"uhlbauer~S., Franz~C., Jonietz~F.,
  Georgii~R., B\"oni~P., Pedersen~B., Schmidt~M., Rosch~A., Pfleiderer~C.,
  Phys. Rev. B, 2010, \textbf{81}, 041203, \doi{10.1103/PhysRevB.81.041203}.

\bibitem{Bauer_PhysRevB.93.235144}
Bauer~A., Garst~M., Pfleiderer~C., Phys. Rev. B, 2016, \textbf{93}, 235144,
  \doi{10.1103/PhysRevB.93.235144}.

\bibitem{chacon2018observation}
Chacon~A., Heinen~L., Halder~M., Bauer~A., Simeth~W., M{\"u}hlbauer~S.,
  Berger~H., Garst~M., Rosch~A., Pfleiderer~C., Nat. Phys., 2018, \textbf{14},
  No.~9, 936--941, \doi{10.1038/s41567-018-0184-y}.

\bibitem{takagi2020particle}
Takagi~R., Yamasaki~Y., Yokouchi~T., Ukleev~V., Yokoyama~Y., Nakao~H.,
  Arima~T., Tokura~Y., Seki~S., Nat. Commun., 2020, \textbf{11}, 5685,
  \doi{10.1038/s41467-020-19480-8}.

\bibitem{kakihana2018giant}
Kakihana~M., Aoki~D., Nakamura~A., Honda~F., Nakashima~M., Amako~Y.,
  Nakamura~S., Sakakibara~T., Hedo~M., Nakama~T., Onuki~Y., J. Phys. Soc. Jpn.,
  2018, \textbf{87}, No.~2, 023701, \doi{10.7566/JPSJ.87.023701}.

\bibitem{kaneko2019unique}
Kaneko~K., Frontzek~M.~D., Matsuda~M., Nakao~A., Munakata~K., Ohhara~T.,
  Kakihana~M., Haga~Y., Hedo~M., Nakama~T., Onuki~Y., J. Phys. Soc. Jpn., 2019,
  \textbf{88}, No.~1, 013702, \doi{10.7566/JPSJ.88.013702}.

\bibitem{tabata2019magnetic}
Tabata~C., Matsumura~T., Nakao~H., Michimura~S., Kakihana~M., Inami~T.,
  Kaneko~K., Hedo~M., Nakama~T., {\=O}nuki~Y., J. Phys. Soc. Jpn., 2019,
  \textbf{88}, No.~9, 093704, \doi{10.7566/JPSJ.88.093704}.

\bibitem{kakihana2019unique}
Kakihana~M., Aoki~D., Nakamura~A., Honda~F., Nakashima~M., Amako~Y.,
  Takeuchi~T., Harima~H., Hedo~M., Nakama~T., Onuki~Y., J. Phys. Soc. Jpn.,
  2019, \textbf{88}, No.~9, 094705, \doi{10.7566/JPSJ.88.094705}.

\bibitem{Mishra_PhysRevB.100.125113}
Mishra~A.~K., Ganesan~V., Phys. Rev. B, 2019, \textbf{100}, 125113,
  \doi{10.1103/PhysRevB.100.125113}.

\bibitem{takeuchi2019magnetic}
Takeuchi~T., Kakihana~M., Hedo~M., Nakama~T., {\=O}nuki~Y., J. Phys. Soc. Jpn.,
  2019, \textbf{88}, No.~5, 053703, \doi{10.7566/JPSJ.88.053703}.

\bibitem{hayami2021field}
Hayami~S., Yambe~R., J. Phys. Soc. Jpn., 2021, \textbf{90}, No.~7, 073705,
  \doi{10.7566/JPSJ.90.073705}.

\bibitem{Matsumura_PhysRevB.109.174437}
Matsumura~T., Tabata~C., Kaneko~K., Nakao~H., Kakihana~M., Hedo~M., Nakama~T.,
  \ifmmode~\bar{O}\else \={O}\fi{}nuki~Y., Phys. Rev. B, 2024, \textbf{109},
  174437, \doi{10.1103/PhysRevB.109.174437}.

\bibitem{Okubo_PhysRevLett.108.017206}
Okubo~T., Chung~S., Kawamura~H., Phys. Rev. Lett., 2012, \textbf{108}, 017206,
  \doi{10.1103/PhysRevLett.108.017206}.

\bibitem{leonov2015multiply}
Leonov~A.~O., Mostovoy~M., Nat. Commun., 2015, \textbf{6}, 8275,
  \doi{10.1038/ncomms9275}.

\bibitem{Hayami_PhysRevB.94.174420}
Hayami~S., Lin~S.-Z., Kamiya~Y., Batista~C.~D., Phys. Rev. B, 2016,
  \textbf{94}, 174420, \doi{10.1103/PhysRevB.94.174420}.

\bibitem{Hayami_PhysRevB.103.224418}
Hayami~S., Phys. Rev. B, 2021, \textbf{103}, 224418,
  \doi{10.1103/PhysRevB.103.224418}.

\bibitem{Hayami_PhysRevB.105.014408}
Hayami~S., Phys. Rev. B, 2022, \textbf{105}, 014408,
  \doi{10.1103/PhysRevB.105.014408}.

\bibitem{lin2024skyrmion}
Lin~S.-Z., Mater. Today Quantum, 2024, \textbf{2}, 100006,
  \doi{10.1016/j.mtquan.2024.100006}.

\bibitem{hayami2024stabilization}
Hayami~S., Yambe~R., Mater. Today Quantum, 2024, \textbf{3}, 100010,
  \doi{10.1016/j.mtquan.2024.100010}.

\bibitem{kawamura2025frustration}
Kawamura~H., J. Phys.: Condens. Matter, 2025, \textbf{37}, No.~18, 183004,
  \doi{10.1088/1361-648X/adbf5b}.

\bibitem{Saha_PhysRevB.60.12162}
Saha~S.~R., Sugawara~H., Matsuda~T.~D., Sato~H., Mallik~R.,
  Sampathkumaran~E.~V., Phys. Rev. B, 1999, \textbf{60}, 12162--12165,
  \doi{10.1103/PhysRevB.60.12162}.

\bibitem{kurumaji2019skyrmion}
Kurumaji~T., Nakajima~T., Hirschberger~M., Kikkawa~A., Yamasaki~Y.,
  Sagayama~H., Nakao~H., Taguchi~Y., Arima~T.-h., Tokura~Y., Science, 2019,
  \textbf{365}, No. 6456, 914--918, \doi{10.1126/science.aau0968}.

\bibitem{sampathkumaran2019report}
Sampathkumaran~E.~V., A report of (topological) hall anomaly two decades ago in
  gd2pdsi3, and its relevance to the history of the field of topological hall
  effect due to magnetic skyrmions, 2019, \doi{10.48550/arXiv.1910.09194},
  \eprint{1910.09194}.

\bibitem{Hirschberger_PhysRevB.101.220401}
Hirschberger~M., Nakajima~T., Kriener~M., Kurumaji~T., Spitz~L., Gao~S.,
  Kikkawa~A., Yamasaki~Y., Sagayama~H., Nakao~H., Ohira-Kawamura~S.,
  Taguchi~Y., Arima~T.-h., Tokura~Y., Phys. Rev. B, 2020, \textbf{101},
  220401(R), \doi{10.1103/PhysRevB.101.220401}.

\bibitem{Kumar_PhysRevB.101.144440}
Kumar~R., Iyer~K.~K., Paulose~P.~L., Sampathkumaran~E.~V., Phys. Rev. B, 2020,
  \textbf{101}, 144440, \doi{10.1103/PhysRevB.101.144440}.

\bibitem{Spachmann_PhysRevB.103.184424}
Spachmann~S., Elghandour~A., Frontzek~M., L\"oser~W., Klingeler~R., Phys. Rev.
  B, 2021, \textbf{103}, 184424, \doi{10.1103/PhysRevB.103.184424}.

\bibitem{Gomilsek_PhysRevLett.134.046702}
Gomil\ifmmode~\check{s}\else \v{s}\fi{}ek~M., Hicken~T.~J., Wilson~M.~N.,
  Franke~K. J.~A., Huddart~B.~M., \ifmmode \check{S}\else
  \v{S}\fi{}tefan\ifmmode \check{c}\else \v{c}\fi{}i\ifmmode~\check{c}\else
  \v{c}\fi{}~A., Holt~S. J.~R., Balakrishnan~G., Mayoh~D.~A., Birch~M.~T.,
  Moody~S.~H., Luetkens~H., Guguchia~Z., Telling~M. T.~F., Baker~P.~J.,
  Clark~S.~J., Lancaster~T., Phys. Rev. Lett., 2025, \textbf{134}, 046702,
  \doi{10.1103/PhysRevLett.134.046702}.

\bibitem{chandragiri2016magnetic}
Chandragiri~V., Iyer~K.~K., Sampathkumaran~E., J. Phys.: Condens. Matter, 2016,
  \textbf{28}, No.~28, 286002, \doi{10.1088/0953-8984/28/28/286002}.

\bibitem{Nakamura_PhysRevB.98.054410}
Nakamura~S., Kabeya~N., Kobayashi~M., Araki~K., Katoh~K., Ochiai~A., Phys. Rev.
  B, 2018, \textbf{98}, 054410, \doi{10.1103/PhysRevB.98.054410}.

\bibitem{hirschberger2019skyrmion}
Hirschberger~M., Nakajima~T., Gao~S., Peng~L., Kikkawa~A., Kurumaji~T.,
  Kriener~M., Yamasaki~Y., Sagayama~H., Nakao~H., Ohishi~K., Kakurai~K.,
  Taguchi~Y., Yu~X., Arima~T.-h., Tokura~Y., Nat. Commun., 2019, \textbf{10},
  No.~1, 5831, \doi{10.1038/s41467-019-13675-4}.

\bibitem{Nakamura_PhysRevB.111.184433}
Nakamura~S., Phys. Rev. B, 2025, \textbf{111}, 184433,
  \doi{10.1103/PhysRevB.111.184433}.

\bibitem{khanh2020nanometric}
Khanh~N.~D., Nakajima~T., Yu~X., Gao~S., Shibata~K., Hirschberger~M.,
  Yamasaki~Y., Sagayama~H., Nakao~H., Peng~L., Nakajima~K., Takagi~R.,
  Arima~T.-h., Tokura~Y., Seki~S., Nat. Nanotechnol., 2020, \textbf{15}, 444,
  \doi{10.1038/s41565-020-0684-7}.

\bibitem{Wood_PhysRevB.107.L180402}
Wood~G. D.~A., Khalyavin~D.~D., Mayoh~D.~A., Bouaziz~J., Hall~A.~E., Holt~S.
  J.~R., Orlandi~F., Manuel~P., Bl\"ugel~S., Staunton~J.~B., Petrenko~O.~A.,
  Lees~M.~R., Balakrishnan~G., Phys. Rev. B, 2023, \textbf{107}, L180402,
  \doi{10.1103/PhysRevB.107.L180402}.

\bibitem{eremeev2023insight}
Eremeev~S.~V., Glazkova~D., Poelchen~G., Kraiker~A., Ali~K., Tarasov~A.~V.,
  Schulz~S., Kliemt~K., Chulkov~E.~V., Stolyarov~V.~S., Ernst~A., Krellner~C.,
  Usachov~D.~{\relax Yu}., Vyalikh~D.~V., Nanoscale Adv., 2023, \textbf{5},
  No.~23, 6678--6687, \doi{10.1039/D3NA00435J}.

\bibitem{Huddart_PhysRevB.111.054440}
Huddart~B.~M., Hern\'andez-Meli\'an~A., Wood~G. D.~A., Mayoh~D.~A.,
  Gomil\ifmmode~\check{s}\else \v{s}\fi{}ek~M., Guguchia~Z., Wang~C.,
  Hicken~T.~J., Blundell~S.~J., Balakrishnan~G., Lancaster~T., Phys. Rev. B,
  2025, \textbf{111}, 054440, \doi{10.1103/PhysRevB.111.054440}.

\bibitem{dong2025pseudogap}
Dong~Y., Kinoshita~Y., Ochi~M., Nakachi~R., Higashinaka~R., Hayami~S., Wan~Y.,
  Arai~Y., Huh~S., Hashimoto~M., Lu~D., Tokunaga~M., Aoki~Y., Matsuda~T.~D.,
  Kondo~T., Science, 2025, \textbf{388}, No. 6747, 624--630,
  \doi{10.1126/science.adj7710}.

\bibitem{Wang_PhysRevLett.124.207201}
Wang~Z., Su~Y., Lin~S.-Z., Batista~C.~D., Phys. Rev. Lett., 2020, \textbf{124},
  207201, \doi{10.1103/PhysRevLett.124.207201}.

\bibitem{Bouaziz_PhysRevLett.128.157206}
Bouaziz~J., Mendive-Tapia~E., Bl\"ugel~S., Staunton~J.~B., Phys. Rev. Lett.,
  2022, \textbf{128}, 157206, \doi{10.1103/PhysRevLett.128.157206}.

\bibitem{wang2023skyrmion}
Wang~Z., Batista~C.~D., SciPost Phys., 2023, \textbf{15}, No.~4, 161,
  \doi{10.21468/SciPostPhys.15.4.161}.

\bibitem{Solenov_PhysRevLett.108.096403}
Solenov~D., Mozyrsky~D., Martin~I., Phys. Rev. Lett., 2012, \textbf{108},
  096403, \doi{10.1103/PhysRevLett.108.096403}.

\bibitem{takagi2018multiple}
Takagi~R., White~J., Hayami~S., Arita~R., Honecker~D., R{\o}nnow~H., Tokura~Y.,
  Seki~S., Sci. Adv., 2018, \textbf{4}, No.~11, eaau3402,
  \doi{10.1126/sciadv.aau3402}.

\bibitem{park2025spin}
Park~P., Cho~W., Kim~C., An~Y., Iida~K., Kajimoto~R., Matin~S., Zhang~S.-S.,
  Batista~C.~D., Park~J.-G., Phys. Rev. X, 2025, \textbf{15}, 031032,
  \doi{10.1103/y9ly-4kld}.

\bibitem{Eto_PhysRevLett.132.226705}
Eto~R., Mochizuki~M., Phys. Rev. Lett., 2024, \textbf{132}, 226705,
  \doi{10.1103/PhysRevLett.132.226705}.

\bibitem{Kato_PhysRevB.104.224405}
Kato~Y., Hayami~S., Motome~Y., Phys. Rev. B, 2021, \textbf{104}, 224405,
  \doi{10.1103/PhysRevB.104.224405}.

\bibitem{Okumura_doi:10.7566/JPSJ.91.093702}
Okumura~S., Hayami~S., Kato~Y., Motome~Y., J. Phys. Soc. Jpn., 2022,
  \textbf{91}, No.~9, 093702, \doi{10.7566/JPSJ.91.093702}.

\bibitem{Hayami_PhysRevB.104.094425}
Hayami~S., Yambe~R., Phys. Rev. B, 2021, \textbf{104}, 094425,
  \doi{10.1103/PhysRevB.104.094425}.

\bibitem{Mohylna_PhysRevB.111.174435}
Mohylna~M., G\'omez~Albarrac\'{\i}n~F.~A., \ifmmode \check{Z}\else
  \v{Z}\fi{}ukovi\ifmmode~\check{c}\else \v{c}\fi{}~M., Rosales~H.~D., Phys.
  Rev. B, 2025, \textbf{111}, 174435, \doi{10.1103/PhysRevB.111.174435}.

\bibitem{Christensen_PhysRevX.8.041022}
Christensen~M.~H., Andersen~B.~M., Kotetes~P., Phys. Rev. X, 2018, \textbf{8},
  041022, \doi{10.1103/PhysRevX.8.041022}.

\bibitem{Hayami_doi:10.7566/JPSJ.89.103702}
Hayami~S., Yambe~R., J. Phys. Soc. Jpn., 2020, \textbf{89}, No.~10, 103702,
  \doi{10.7566/JPSJ.89.103702}.

\bibitem{Utesov_PhysRevB.103.064414}
Utesov~O.~I., Phys. Rev. B, 2021, \textbf{103}, 064414,
  \doi{10.1103/PhysRevB.103.064414}.

\bibitem{Wang_PhysRevB.103.104408}
Wang~Z., Su~Y., Lin~S.-Z., Batista~C.~D., Phys. Rev. B, 2021, \textbf{103},
  104408, \doi{10.1103/PhysRevB.103.104408}.

\bibitem{Hayami_PhysRevB.105.174437}
Hayami~S., Phys. Rev. B, 2022, \textbf{105}, 174437,
  \doi{10.1103/PhysRevB.105.174437}.

\bibitem{Hayami_PhysRevB.105.104428}
Hayami~S., Yambe~R., Phys. Rev. B, 2022, \textbf{105}, 104428,
  \doi{10.1103/PhysRevB.105.104428}.

\bibitem{utesov2025thermodynamic}
Utesov~O.~I., Budylev~D.~P., Phys. Rev. B, 2025, \textbf{112}, 224432,
  \doi{10.1103/k62j-1lyz}.

\bibitem{hayami2022multiple}
Hayami~S., J. Phys. Soc. Jpn., 2022, \textbf{91}, No.~2, 023705,
  \doi{10.7566/JPSJ.91.023705}.

\bibitem{Yambe_PhysRevB.106.174437}
Yambe~R., Hayami~S., Phys. Rev. B, 2022, \textbf{106}, 174437,
  \doi{10.1103/PhysRevB.106.174437}.

\bibitem{hayami2020multiple}
Hayami~S., J. Magn. Magn. Mater., 2020, \textbf{513}, 167181,
  \doi{10.1016/j.jmmm.2020.167181}.

\bibitem{hayami2023widely}
Hayami~S., Kato~Y., J. Magn. Magn. Mater., 2023, \textbf{571}, 170547,
  \doi{10.1016/j.jmmm.2023.170547}.

\bibitem{Ruderman}
Ruderman~M.~A., Kittel~C., Phys. Rev., 1954, \textbf{96}, 99--102,
  \doi{10.1103/PhysRev.96.99}.

\bibitem{Kasuya}
Kasuya~T., Prog. Theor. Phys., 1956, \textbf{16}, No.~1, 45--57,
  \doi{10.1143/PTP.16.45}.

\bibitem{Yosida1957}
Yosida~K., Phys. Rev., 1957, \textbf{106}, 893--898,
  \doi{10.1103/PhysRev.106.893}.

\bibitem{yoshimori1959new}
Yoshimori~A., J. Phys. Soc. Jpn., 1959, \textbf{14}, No.~6, 807--821.

\bibitem{Kaplan_PhysRev.124.329}
Kaplan~T.~A., Phys. Rev., 1961, \textbf{124}, 329--339,
  \doi{10.1103/PhysRev.124.329}.

\bibitem{Elliott_PhysRev.124.346}
Elliott~R.~J., Phys. Rev., 1961, \textbf{124}, 346--353,
  \doi{10.1103/PhysRev.124.346}.

\bibitem{day1981neutron}
Day~P., Moore~M.~W., Wilkinson~C., Ziebeck~K. R.~A., J. Phys. C: Solid State
  Phys., 1981, \textbf{14}, No.~23, 3423, \doi{10.1088/0022-3719/14/23/018}.

\bibitem{regnault1982inelastic}
Regnault~L., Rossat-Mignod~J., Adam~A., Billerey~D., Terrier~C., J. Phys.
  France, 1982, \textbf{43}, No.~8, 1283--1290,
  \doi{10.1051/jphys:019820043080128300}.

\bibitem{nakatsuji2005spin}
Nakatsuji~S., Nambu~Y., Tonomura~H., Sakai~O., Jonas~S., Broholm~C.,
  Tsunetsugu~H., Qiu~Y., Maeno~Y., Science, 2005, \textbf{309}, No. 5741,
  1697--1700, \doi{10.1126/science.1114727}.

\bibitem{yambe2021skyrmion}
Yambe~R., Hayami~S., Sci. Rep., 2021, \textbf{11}, 11184,
  \doi{10.1038/s41598-021-90308-1}.

\bibitem{seo2021spin}
Seo~S., Hayami~S., Su~Y., Thomas~S.~M., Ronning~F., Bauer~E.~D.,
  Thompson~J.~D., Lin~S.-Z., Rosa~P.~F., Commun. Phys., 2021, \textbf{4},
  No.~1, 58, \doi{10.1038/s42005-021-00558-8}.

\bibitem{singh2023transition}
Singh~D., Fujishiro~Y., Hayami~S., Moody~S.~H., Nomoto~T., Baral~P.~R.,
  Ukleev~V., Cubitt~R., Steinke~N.-J., Gawryluk~D.~J., Pomjakushina~E.,
  {\=O}nuki~Y., Arita~R., Tokura~Y., Kanazawa~N., White~J.~S., Nat. Commun.,
  2023, \textbf{14}, 8050, \doi{10.1038/s41467-023-43814-x}.

\bibitem{neves2026cascade}
Neves~P.~M., Kurumaji~T., Wakefield~J.~P., Ip~C. I.~J., Cubitt~R., Hayami~S.,
  White~J.~S., Checkelsky~J.~G., ACS Nano, 2026, \textbf{20}, No.~19,
  14029--14038, \doi{https://doi.org/10.1021/acsnano.5c18732}.

\bibitem{neves2026general}
Neves~P.~M., Kurumaji~T., Wakefield~J.~P., Hiess~A., Steffens~P., Qureshi~N.,
  Cubitt~R., DeBeer-Schmitt~L.~M., Palmstrom~J.~C., Hayami~S., Bartkowiak~M.,
  Zolliker~M., White~J.~S., Checkelsky~J.~G., Phys. Rev. X, 2026, \textbf{16},
  021054, \doi{10.1103/bjq3-py7l}.

\bibitem{Hayami_PhysRevResearch.3.043158}
Hayami~S., Yambe~R., Phys. Rev. Res., 2021, \textbf{3}, 043158,
  \doi{10.1103/PhysRevResearch.3.043158}.

\end{thebibliography}

\newpage
\ukrainianpart
\title[Конкуренція між однорідним феромагнетизмом та нестійкістю зі скінченним хвильовим вектором у скірміонних кристалах на квадратній ґратці]%
{Конкуренція між однорідним феромагнетизмом та нестійкістю зі скінченним хвильовим вектором у скірміонних кристалах на квадратній ґратці}
\author[С. Хаямі]{С. Хаямі
}
\address{Вища школа наук, Університет Хоккайдо, Саппоро 060-0810, Японія
}

\makeukrtitle
\begin{abstract}
Ми досліджуємо стани скірміонного кристалу на квадратній ґратці в центросиметричній спіновій моделі з магнітними взаємодіями, що залежать від імпульсу, включаючи однорідну феромагнітну взаємодію, анізотропні взаємодії ``легка вісь'' на двох ортогональних скінченних хвильових векторах, взаємодії хвильових векторів вищих гармонік та зовнішнє магнітне поле.
Використовуючи розрахунки моделювання відпалу, побудовано низькотемпературну магнітну фазову діаграму як функцію магнітного поля та однорідної феромагнітної взаємодії.
Виявлено, що кристалічна фаза скірміона стабілізується в проміжних полях, коли однорідна феромагнітна взаємодія слабка, демонструючи некомпланарну подвійну $Q$ спінову текстуру, скінченну скалярну спінову хіральність та компоненти вищих гармонік. Збільшення однорідної феромагнітної взаємодії пригнічує та, зрештою, знищує кристалічну фазу скірміона через прямі переходи першого порядку до повністю поляризованого стану, тоді як її зменшення розширює область стабільності скірміона та стабілізує конічну спіральну фазу з одним $Q$ у сильних полях. Ці результати уточнюють механізми, що регулюють формування скірміонів квадратної ґратки в центросиметричних магнетиках.
	\keywords скірміонний кристал, квадратна ґратка, центросиметричні магнетики, феромагнітна взаємодія, магнітна анізотропія, взаємодія хвильового вектора вищих гармонік
\end{abstract}

  \end{document}